\documentclass[notitlepage,preprintnumbers,prd,nofootinbib]{revtex4-1}

\usepackage{amsmath,amssymb,slashed}
\usepackage{hyperref}
\usepackage{url}
\usepackage{breakurl}
\usepackage{graphicx}
\usepackage{subfigure}

\begin{document}


\title{A Composite Model for the 750 GeV Diphoton Excess}

\author{Keisuke Harigaya and Yasunori Nomura}
\affiliation{Berkeley Center for Theoretical Physics, Department of Physics, 
  University of California, Berkeley, CA 94720}
\affiliation{Theoretical Physics Group, Lawrence Berkeley National 
  Laboratory, Berkeley, CA 94720}

\begin{abstract}
We study a simple model in which the recently reported $750~{\rm GeV}$ 
diphoton excess arises from a composite pseudo~Nambu-Goldstone 
boson---hidden pion---produced by gluon fusion and decaying into 
two photons.  The model only introduces an extra hidden gauge 
group at the TeV scale with a vectorlike quark in the bifundamental 
representation of the hidden and standard model gauge groups. 
We calculate the masses of all the hidden pions and analyze their 
experimental signatures and constraints.  We find that two colored 
hidden pions must be near the current experimental limits, and hence 
are probed in the near future.  We study physics of would-be stable 
particles---the composite states that do not decay purely by the 
hidden and standard model gauge dynamics---in detail, including 
constraints from cosmology.  We discuss possible theoretical structures 
above the TeV scale, e.g.\ conformal dynamics and supersymmetry, 
and their phenomenological implications.  We also discuss an extension 
of the minimal model in which there is an extra hidden quark that 
is singlet under the standard model and has a mass smaller than 
the hidden dynamical scale.  This provides two standard model singlet 
hidden pions that can both be viewed as diphoton/diboson resonances 
produced by gluon fusion.  We discuss several scenarios in which these 
(and other) resonances can be used to explain various excesses seen 
in the LHC data.
\end{abstract}

\maketitle

\section{Introduction}
\label{sec:intro}

If the recently announced diphoton excess at $\simeq 750~{\rm 
GeV}$~\cite{ATLAS,CMS} remains as a true signal, it indicates a 
long-awaited discovery of new physics beyond the standard model. 
In a recent paper~\cite{Harigaya:2015ezk}, we have proposed that 
this excess may result from a composite spin-$0$ particle decaying 
into a two-photon final state.  Among the possibilities discussed, 
here we concentrate on the case in which the particle is a composite 
pseudo~Nambu-Goldstone boson associated with new strong dynamics at 
the TeV scale, which is singly produced by gluon fusion and decays 
into two photons.  For works that have appeared around the same time 
and discussed similar models to those in~\cite{Harigaya:2015ezk}, 
see~\cite{Nakai:2015ptz}; other related works include~\cite{Low:2015qep}. 
A class of theories involving similar dynamics with vectorlike matter 
charged under both hidden and standard model gauge groups was studied 
in~\cite{Kilic:2009mi,Kilic:2010et}.  The possibility of obtaining 
standard model dibosons from a composite scalar particle was utilized 
in a different context in~\cite{Chiang:2015lqa,Cacciapaglia:2015nga}.

In this paper, we study a simple model presented in 
Ref.~\cite{Harigaya:2015ezk} which introduces an extra gauge group 
$G_H = SU(N)$ at the TeV scale, in addition to the standard model gauge 
group $G_{\rm SM} = SU(3)_C \times SU(2)_L \times U(1)_Y$, with extra 
matter---hidden quarks---in the vectorlike bifundamental representation 
of $G_H$ and $SU(5) \supset G_{\rm SM}$.  Here, $SU(5)$ is used only as 
a simple mnemonic; it does not mean that the three factors of $G_{\rm SM}$ 
are unified at the TeV scale.  In this model, the diphoton resonance is 
one of the pseudo~Nambu-Goldstone bosons---hidden pions---which is neutral 
under $G_{\rm SM}$ and has the mass of $\simeq 750~{\rm GeV}$.  We consider 
this particular model here because it is theoretically simple and leads 
to predictions that can be tested in the near future.  The only free 
parameters of the model, beyond the size $N$ of $G_H$, are the dynamical 
scale of $G_H$ and two (in general complex) masses for the hidden quarks, 
out of which two numbers are fixed by the mass and diphoton rate of 
the resonance.  We calculate all the masses of the hidden pions, which 
for a given $N$ depend only on a single free parameter:\ the ratio of 
the absolute values of the two hidden quark masses.  We find that two 
colored hidden pions must be near the current experimental limits (unless 
$N$ is unreasonably large); they respectively lead to narrow dijet, 
$Z$-jet, and $\gamma$-jet resonances below $\sim 1.6~{\rm TeV}$ and to 
heavy stable charged and neutral hadrons (or leptoquark-type resonances) 
below $\sim 1.2~{\rm TeV}$.  We also discuss other, higher resonances 
in the model, which are expected to be in the multi-TeV region, as well 
as the effect of possible $CP$ violation in the $G_H$ sector on the 
standard model physics.  Furthermore, we investigate what the structure 
of the model above the TeV region can be.  This includes the possibility 
of (a part of) the theory being conformal and/or supersymmetric.  This 
affects collider signatures and cosmological implications of one of 
the colored hidden pions:\ the one that does not decay through $G_H$ 
or standard model gauge dynamics.  We also perform detailed analyses 
of cosmology of hidden baryons.

We then discuss an extension of the minimal model which has an extra 
hidden quark that is singlet under the standard model gauge group.  One 
of the salient features of this model is that there are two diphoton 
(diboson) resonances in the hidden pion sector, which can both be produced 
by gluon fusion and decay into two electroweak gauge bosons.  We consider 
several scenarios associated with these and other related resonances. 
Representative ones are:\ (i) the two resonances correspond to two 
diphoton ``excesses'' seen in the ATLAS data at $\simeq 750~{\rm GeV}$ 
and $\simeq 1.6~{\rm TeV}$~\cite{ATLAS} (although the latter is much 
weaker than the former); (ii) the two resonances are both near $\simeq 
750~{\rm GeV}$ with a mass difference of 10s of GeV, explaining a slight 
preference to a wide width in the ATLAS data; (iii) the two resonances 
are at $\simeq 750~{\rm GeV}$ and $\simeq 2~{\rm TeV}$, responsible for 
the $750~{\rm GeV}$ diphoton excess~\cite{ATLAS,CMS} and the $2~{\rm TeV}$ 
diboson excess~\cite{Aad:2015owa}, respectively.  We calculate the masses 
of the hidden pions in the model and discuss their phenomenology.  We 
find that the masses of the hidden pions can be larger than the case 
without the singlet hidden quark; in particular, the leptoquark type 
hidden pion can be as heavy as $\sim 1.5~{\rm TeV}$, depending on 
scenarios.  We discuss physics of hidden pions and hidden baryons 
that decay only through interactions beyond the $G_H$ and standard 
model gauge dynamics.  We find that cosmological constraints on this 
model are weaker than those in the model without the extra hidden quark.

The organization of this paper is as follows.  In Section~\ref{sec:model}, 
we consider the minimal model and its phenomenology at the TeV scale. 
We calculate all the hidden pion masses and discuss their signatures 
and current constraints.  We also discuss particles with higher masses, 
in particular the hidden $\eta'$ meson and spin-1 resonances.  In 
Section~\ref{sec:higher-e}, we study physics above the TeV scale, 
especially its implications for collider physics and cosmology.  The 
hidden pion that is stable under the $G_H$ and $G_{\rm SM}$ gauge dynamics 
as well as low-lying hidden baryons are studied in detail.  We discuss 
the possibility that the $G_H$ sector is conformal and/or that the theory 
is supersymmetric above the TeV scale.  In Section~\ref{sec:6-flavor}, 
we study an extension of the model in which there is an extra hidden quark 
that is singlet under $G_{\rm SM}$ and has a mass smaller than $\Lambda$. 
We discuss possible signals of two $G_{\rm SM}$-singlet hidden pions 
which can be viewed as diboson resonances produced by gluon fusion. 
Section~\ref{sec:discuss} is devoted to final discussion.  In the Appendix, 
we analyze the effect of possible $CP$ violation in the $G_H$ sector on 
the standard model physics.

\section{Model at the TeV Scale}
\label{sec:model}

The model at the TeV scale is given by a hidden gauge group $G_H = SU(N)$, 
with the dynamical scale (the mass scale of generic low-lying resonances) 
$\Lambda$, and hidden quarks charged under both $G_H$ and the standard 
model gauge groups as shown in Table~\ref{tab:model}.%
\footnote{Throughout the paper, we adopt the hypercharge normalization 
 such that the standard model left-handed Weyl fermions have 
 $(q,u,d,l,e) = (1/6,-2/3,1/3,-1/2,1)$.}
\begin{table}[t]
\begin{center}
\begin{tabular}{c|cccc}
   & $G_H = SU(N)$ & $SU(3)_C$ & $SU(2)_L$ & $U(1)_Y$ \\ \hline
 $\Psi_D$       &       $\Box$ & $\bar{\Box}$ & ${\bf 1}$ &  $1/3$ \\
 $\Psi_L$       &       $\Box$ &    ${\bf 1}$ &    $\Box$ & $-1/2$ \\
 $\bar{\Psi}_D$ & $\bar{\Box}$ &       $\Box$ & ${\bf 1}$ & $-1/3$ \\
 $\bar{\Psi}_L$ & $\bar{\Box}$ &    ${\bf 1}$ &    $\Box$ &  $1/2$ 
\end{tabular}
\end{center}
\caption{Charge assignment of hidden quarks under the hidden and standard 
 model gauge groups. Here, $\Psi_{D,L}$ and $\bar{\Psi}_{D,L}$ are 
 left-handed Weyl spinors.}
\label{tab:model}
\end{table}
The hidden quarks have mass terms
\begin{equation}
  {\cal L} = -m_D \Psi_D \bar{\Psi}_D - m_L \Psi_L \bar{\Psi}_L + {\rm h.c.},
\label{eq:L_mass}
\end{equation}
where we take $m_{D,L} > 0$, which does not lead to a loss of generality 
if we keep all the phases in the other part of the theory.  These masses 
are assumed to be sufficiently smaller than the dynamical scale, $m_{D,L} 
\ll \Lambda$, so that $\Psi_{D,L}$ and $\bar{\Psi}_{D,L}$ can be regarded 
as light quarks from the point of view of the $G_H$ dynamics.  Note 
that the charge assignment of the hidden quarks is such that they are 
a vectorlike fermion in the bifundamental representation of $G_H$ and 
$SU(5) \supset G_{\rm SM}$.  The model therefore preserves gauge coupling 
unification at the level of the standard model; this is significant 
especially given the possible threshold corrections around the TeV 
and unification scales (see, e.g.,~\cite{Hall:2009nd}).  The unification 
of the couplings becomes even better if we introduce supersymmetry 
near the TeV scale (see Section~\ref{sec:higher-e}).

\subsection{Hidden Pion for the 750 GeV Diphoton Excess}
\label{subsec:750-GeV}

The strong $G_H$ dynamics makes the hidden quarks condensate
\begin{equation}
  \langle \Psi_D \bar{\Psi}_D + \Psi_D^\dagger \bar{\Psi}_D^\dagger \rangle 
  \approx \langle \Psi_L \bar{\Psi}_L 
    + \Psi_L^\dagger \bar{\Psi}_L^\dagger \rangle 
  \equiv - c.
\label{eq:PsiPsi-cond}
\end{equation}
These condensations do not break the standard model gauge groups, since 
the hidden quark quantum numbers under these gauge groups are vectorlike 
with respect to $G_H$~\cite{Vafa:1983tf}.  The spectrum below $\Lambda$ 
then consists of hidden pions, arising from spontaneous breaking of 
the approximate $SU(5)_A$ axial flavor symmetry:
\begin{equation}
  \psi({\bf Adj}, {\bf 1}, 0), \qquad
  \chi\Bigl(\Box, \Box, -\frac{5}{6}\Bigr), \qquad
  \varphi({\bf 1}, {\bf Adj}, 0), \qquad
  \phi({\bf 1}, {\bf 1}, 0),
\label{eq:pions_unif}
\end{equation}
where $\psi$, $\varphi$, and $\phi$ are real scalars while $\chi$ is a 
complex scalar, and the quantum numbers represent those under $SU(3)_C 
\times SU(2)_L \times U(1)_Y$.  The masses of these particles are given 
by~\cite{Weinberg:1996kr}
\begin{align}
  m_\psi^2 &= 2 m_D \frac{c}{f^2} 
    + 3 \Delta_C,
\label{eq:m_psi}\\
  m_\chi^2 &= (m_D + m_L) \frac{c}{f^2} 
    + \frac{4}{3} \Delta_C + \frac{3}{4} \Delta_L + \frac{5}{12} \Delta_Y,
\label{eq:m_chi}\\
  m_\varphi^2 &= 2 m_L \frac{c}{f^2} 
    + 2 \Delta_L,
\label{eq:m_varphi}\\
  m_\phi^2 &= \frac{4 m_D + 6 m_L}{5} \frac{c}{f^2}.
\label{eq:m_phi}
\end{align}
Here, $f$ is the decay constant,%
\footnote{Our definition of the decay constant, $f$, is a factor of $2$ 
 different from that in Ref.~\cite{Weinberg:1996kr}:\ $f = F/2$.}
and $\Delta_{C,L,Y}$ are contributions from standard model gauge loops, 
of order
\begin{equation}
  \Delta_C \simeq \frac{3 g_3^2}{16\pi^2} \Lambda^2,
\qquad
  \Delta_L \simeq \frac{3 g_2^2}{16\pi^2} \Lambda^2,
\qquad
  \Delta_Y \simeq \frac{3 g_1^2}{16\pi^2} \Lambda^2,
\label{eq:Delta_CLY}
\end{equation}
where $g_3$, $g_2$, and $g_1$ are the gauge couplings of $SU(3)_C$, 
$SU(2)_L$, and $U(1)_Y$, respectively, with $g_1$ in the $SU(5)$ 
normalization.  Using naive dimensional analysis~\cite{Manohar:1983md}, 
we can estimate the quark bilinear condensate and the decay constant as
\begin{equation}
  c \approx \frac{N}{16\pi^2} \Lambda^3,
\qquad
  f \approx \frac{\sqrt{N}}{4\pi} \Lambda,
\label{eq:scales}
\end{equation}
where we have assumed $N \gtrsim 5$, i.e.\ the number of color is not much 
smaller than that of flavor in the $G_H$ gauge theory.  For $N \ll 5$, 
we might instead have $c \approx (5/16\pi^2) \Lambda^3$ and $f \approx 
(\sqrt{5}/4\pi) \Lambda$, but below we use Eq.~(\ref{eq:scales}) even 
in this case because the resulting differences are insignificant for 
our results.

The couplings of the hidden pions with the standard model gauge fields 
are determined by chiral anomalies and given by
\begin{align}
  {\cal L} =& \frac{N g_3^2}{64\pi^2 f} d^{abc} \psi^a 
    \epsilon^{\mu\nu\rho\sigma} G^b_{\mu\nu} G^c_{\rho\sigma} 
   + \frac{N g_3 g_1}{16\sqrt{15}\pi^2 f} \psi^a 
    \epsilon^{\mu\nu\rho\sigma} G^a_{\mu\nu} B_{\rho\sigma} 
\nonumber\\
  & - \frac{3 N g_2 g_1}{32\sqrt{15}\pi^2 f} \varphi^\alpha 
    \epsilon^{\mu\nu\rho\sigma} W^\alpha_{\mu\nu} B_{\rho\sigma} 
\nonumber\\
  & + \frac{N g_3^2}{32\sqrt{15}\pi^2 f} \phi\, 
    \epsilon^{\mu\nu\rho\sigma} G^a_{\mu\nu} G^a_{\rho\sigma} 
  - \frac{3 N g_2^2}{64\sqrt{15}\pi^2 f} \phi\, 
    \epsilon^{\mu\nu\rho\sigma} W^\alpha_{\mu\nu} W^\alpha_{\rho\sigma} 
  - \frac{N g_1^2}{64\sqrt{15}\pi^2 f} \phi\, 
    \epsilon^{\mu\nu\rho\sigma} B_{\mu\nu} B_{\rho\sigma},
\label{eq:pion-couplings}
\end{align}
where $a,b,c = 1,\cdots,8$ and $\alpha=1,2,3$ are $SU(3)_C$ and $SU(2)_L$ 
adjoint indices, respectively, and $d^{abc} \equiv 2 {\rm tr}[t^a \{ t^b, 
t^c \}]$ with $t^a$ being half of the Gell-Mann matrices.  We assume that 
the $\phi$ particle produced by gluon fusion and decaying to a diphoton 
is responsible for the observed excess~\cite{Harigaya:2015ezk}, so we take
\begin{equation}
  m_\phi \simeq 750~{\rm GeV}.
\label{eq:m_phi-750}
\end{equation}
The decay of $\phi$ occurs through interactions in 
Eq.~(\ref{eq:pion-couplings}) and leads to standard model gauge bosons. 
The diphoton rate at $\sqrt{s} = 13~{\rm TeV}$ is given by
\begin{equation}
  \sigma(pp \rightarrow \phi \rightarrow \gamma\gamma) 
  \simeq 7.8~{\rm fb} \left( \frac{N}{5} \frac{500~{\rm GeV}}{f} \right)^2.
\label{eq:diphoton-rate}
\end{equation}
Here, we have used the NNPDF~3.0 parton distribution 
function~\cite{Ball:2014uwa} and determined the overall normalization 
(the QCD $K$ factor) such that it reproduces the production cross 
section of a standard~model-like Higgs boson of mass $750~{\rm GeV}$ 
at $\sqrt{s}=14~{\rm TeV}$~\cite{LHC-Higgs}.  Since the observed excess 
corresponds to $\sigma(pp \rightarrow \phi \rightarrow \gamma\gamma) 
\simeq (6 \pm 2)~{\rm fb}$~\cite{Buttazzo:2015txu}, this gives
\begin{equation}
  f \simeq 570~{\rm GeV}\, \frac{N}{5} \sqrt{\frac{6~{\rm fb}} 
    {\sigma(pp \rightarrow \phi \rightarrow \gamma\gamma)}}.
\label{eq:f}
\end{equation}
With this value of $f$, the upper limits from searches in the $8~{\rm TeV}$ 
data~\cite{Aad:2015mna} are evaded.

The ratios of branching fractions to various $\phi$ decay modes are 
given by
\begin{gather}
\begin{aligned}
  & \frac{B_{\phi \rightarrow gg}}{B_{\phi \rightarrow \gamma\gamma}} 
  = 8 \left( \frac{6 g_3^2}{14 e^2} \right)^2 
  \simeq 200,
\qquad
 && \frac{B_{\phi \rightarrow WW}}{B_{\phi \rightarrow \gamma\gamma}} 
  = 2 \left( \frac{9}{14 \sin^2\!\theta_W} \right)^2 
  \simeq 15,
\\
  & \frac{B_{\phi \rightarrow ZZ}}{B_{\phi \rightarrow \gamma\gamma}} 
  = \left( \frac{9 + 5 \tan^4\!\theta_W}{14 \tan^2\!\theta_W} \right)^2 
  \simeq 5,
\qquad
 && \frac{B_{\phi \rightarrow Z\gamma}}{B_{\phi \rightarrow \gamma\gamma}} 
  = 2 \left( \frac{9 - 5 \tan^2\!\theta_W}{14 \tan\theta_W} \right)^2 
  \simeq 2,
\label{eq:phi_ZG-GG}
\end{aligned}
\end{gather}
where $e$ and $\theta_W$ are the electromagnetic coupling and the 
Weinberg angle, respectively, and we have ignored the phase space 
factors.  These values are consistent with the constraints from 
searches of high-mass diboson and dijet resonances in the $8~{\rm TeV}$ 
data~\cite{Harigaya:2015ezk,Knapen:2015dap}.  Observing these decay 
modes in the $13~{\rm TeV}$ run with the predicted rates would provide 
an important test of the model.

\subsection{Other Hidden Pions}
\label{subsec:pions}

The identification of $\phi$ as the $750~{\rm GeV}$ diphoton resonance 
leads, through Eq.~(\ref{eq:m_phi}), to
\begin{equation}
  \frac{2 m_D + 3 m_L}{5} \sim 90~{\rm GeV} \sqrt{\frac{5}{N}} \sqrt{
    \frac{\sigma(pp \rightarrow \phi \rightarrow \gamma\gamma)}{6~{\rm fb}}},
\label{eq:m-1}
\end{equation}
where we have used Eqs.~(\ref{eq:scales},~\ref{eq:f}).  This, 
however, leaves the ratio $r \equiv m_D/m_L$ undetermined. 
With $\Lambda = 3.2~{\rm TeV} \sqrt{N/5}$, which is motivated 
by Eqs.~(\ref{eq:scales},~\ref{eq:f}), the masses of the other hidden 
pions are determined by Eqs.~(\ref{eq:m_psi}~--~\ref{eq:m_varphi}) 
in terms of $r$:
\begin{align}
  m_\psi^2 &\simeq \frac{5r}{2r+3} (750~{\rm GeV})^2 
    + \frac{N}{5} (760~{\rm GeV})^2,
\label{eq:m_psi-2}\\
  m_\chi^2 &\simeq \frac{5r+5}{4r+6} (750~{\rm GeV})^2 
    + \frac{N}{5} (580~{\rm GeV})^2,
\label{eq:m_chi-2}\\
  m_\varphi^2 &\simeq \frac{5}{2r+3} (750~{\rm GeV})^2 
    + \frac{N}{5} (400~{\rm GeV})^2.
\label{eq:m_varphi-2}
\end{align}
Here, in the second terms we have used Eq.~(\ref{eq:Delta_CLY}) with 
unit coefficients, but we do not expect that using the true coefficients 
(which are not known in general) make a significant difference.
\begin{figure}[t]
\centering
  \subfigure{\includegraphics[clip,width=.49\textwidth]{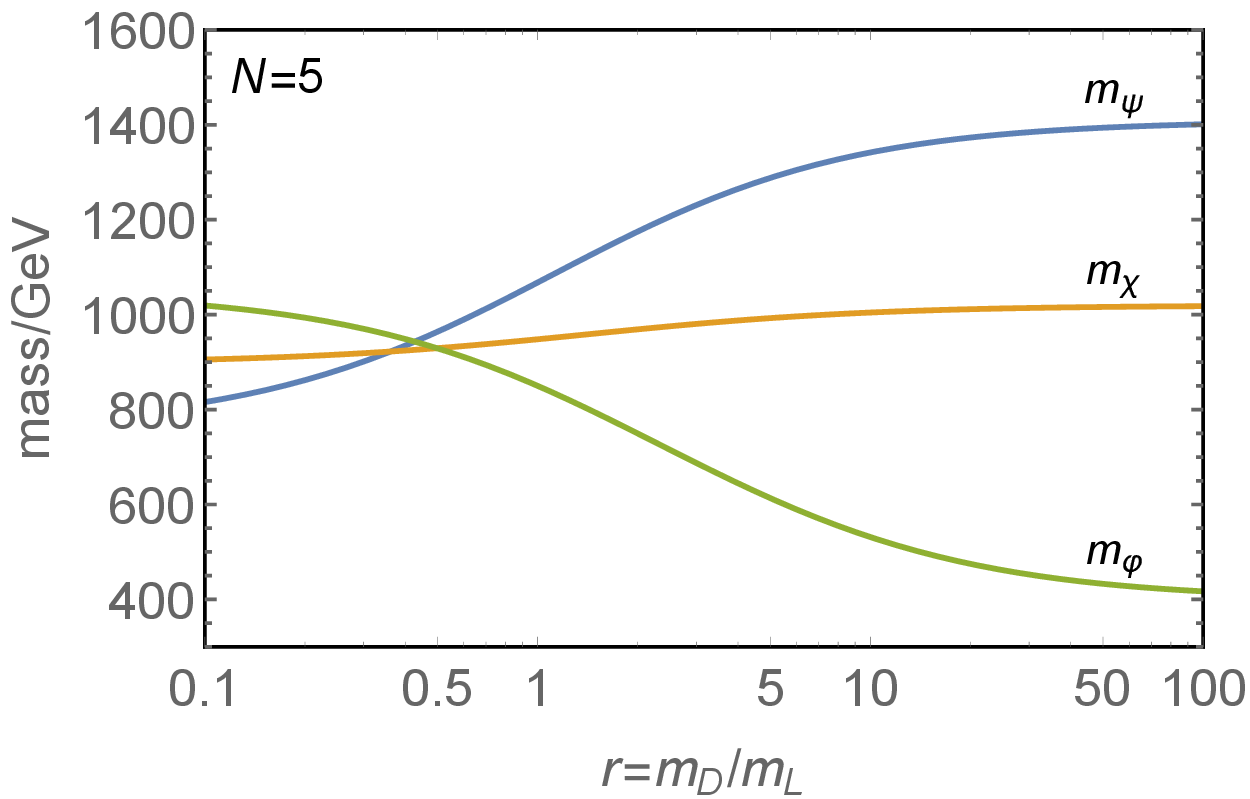}}
\hspace{3mm}
  \subfigure{\includegraphics[clip,width=.48\textwidth]{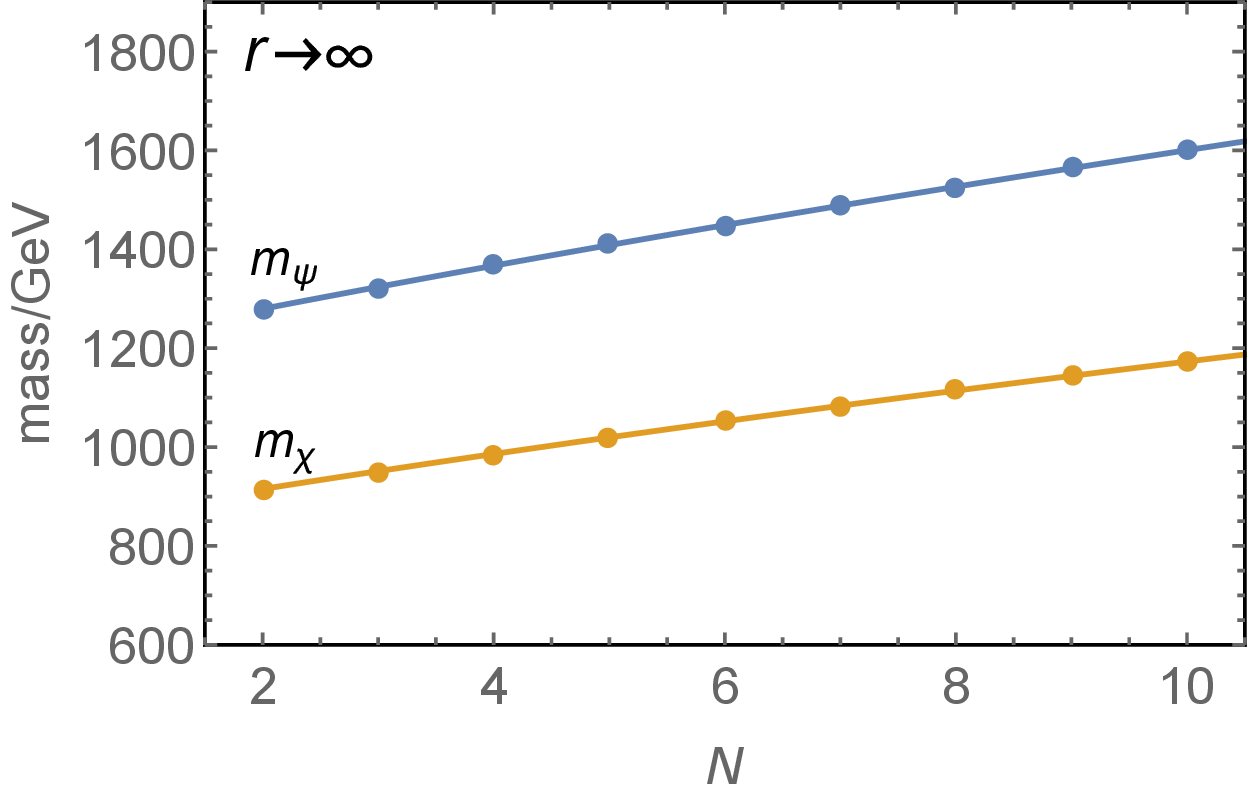}}
\caption{The masses of hidden pions $\psi$, $\chi$, and $\varphi$ 
 for $m_\phi = 750~{\rm GeV}$ as functions of $r = m_D/m_L$ for 
 $N=5$ (left) and the maximal values of the $\psi$ and $\chi$ masses, 
 $m_\psi|_{r \rightarrow \infty}$ and $m_\chi|_{r \rightarrow \infty}$, 
 as functions of $N$ (right).  Here, we have taken $\Lambda = 
 3.2~{\rm TeV} \sqrt{N/5}$, motivated by Eqs.~(\ref{eq:scales},~\ref{eq:f}), 
 and used Eq.~(\ref{eq:Delta_CLY}) with unit coefficients.}
\label{fig:pion-masses}
\end{figure}
In the left panel of Fig.~\ref{fig:pion-masses}, we plot these masses 
for $N=5$ as functions of $r$.  In the right panel, we plot the maximal 
values of the $\psi$ and $\chi$ masses, $m_\psi|_{r \rightarrow \infty}$ 
and $m_\chi|_{r \rightarrow \infty}$, as functions of $N$.  We find 
that these particles are at
\begin{equation}
  m_\psi \lesssim 1.6~{\rm TeV},
\qquad
  m_\chi \lesssim 1.2~{\rm TeV},
\label{eq:psi-chi}
\end{equation}
unless $N$ is very large, $N \geq 10$.  We note that if $m_D$ and 
$m_L$ are unified at a conventional unification scale (around 
$10^{14}\mbox{--}10^{17}~{\rm GeV}$), then their ratio at the TeV 
scale is in the range $r \simeq 1.5~\mbox{--}~3$, with the precise 
value depending on the structure of the theory above the TeV scale.

\begin{figure}[t]
\centering
  \subfigure{\includegraphics[clip,width=.49\textwidth]{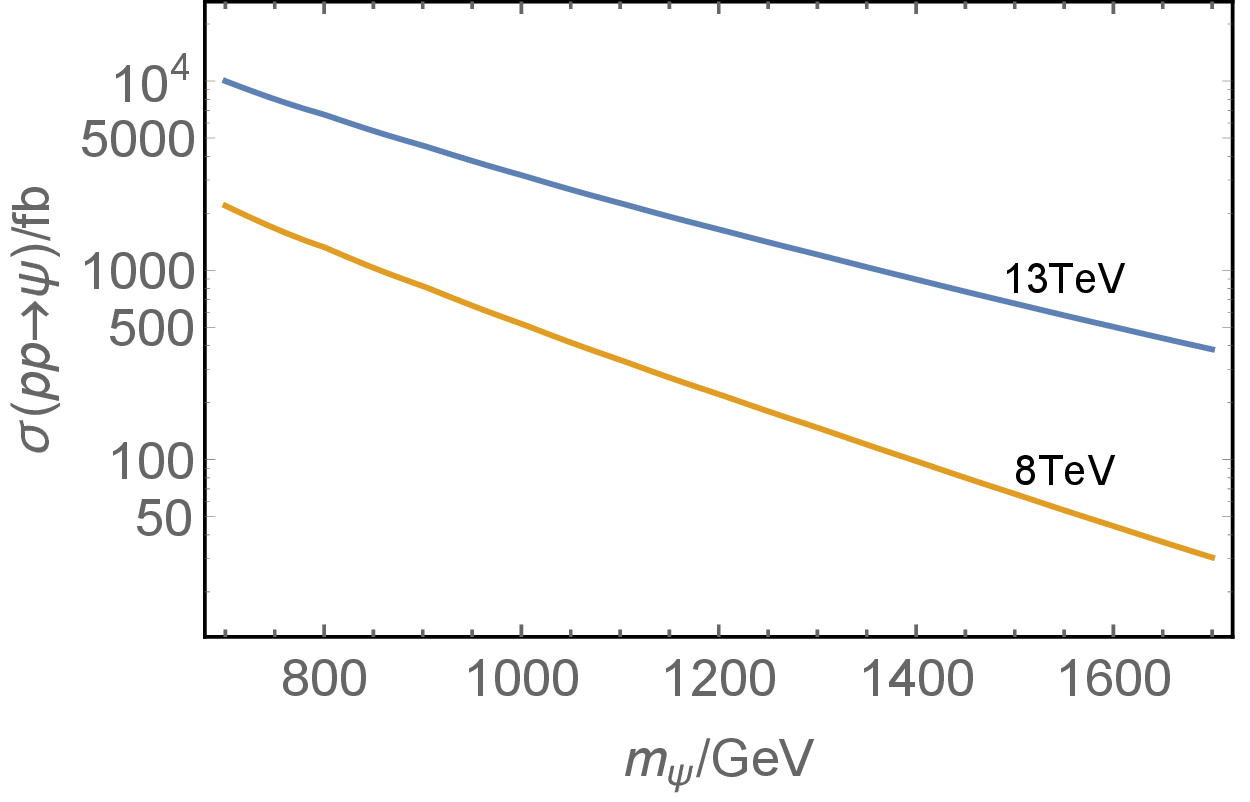}}
\hspace{3mm}
  \subfigure{\includegraphics[clip,width=.48\textwidth]{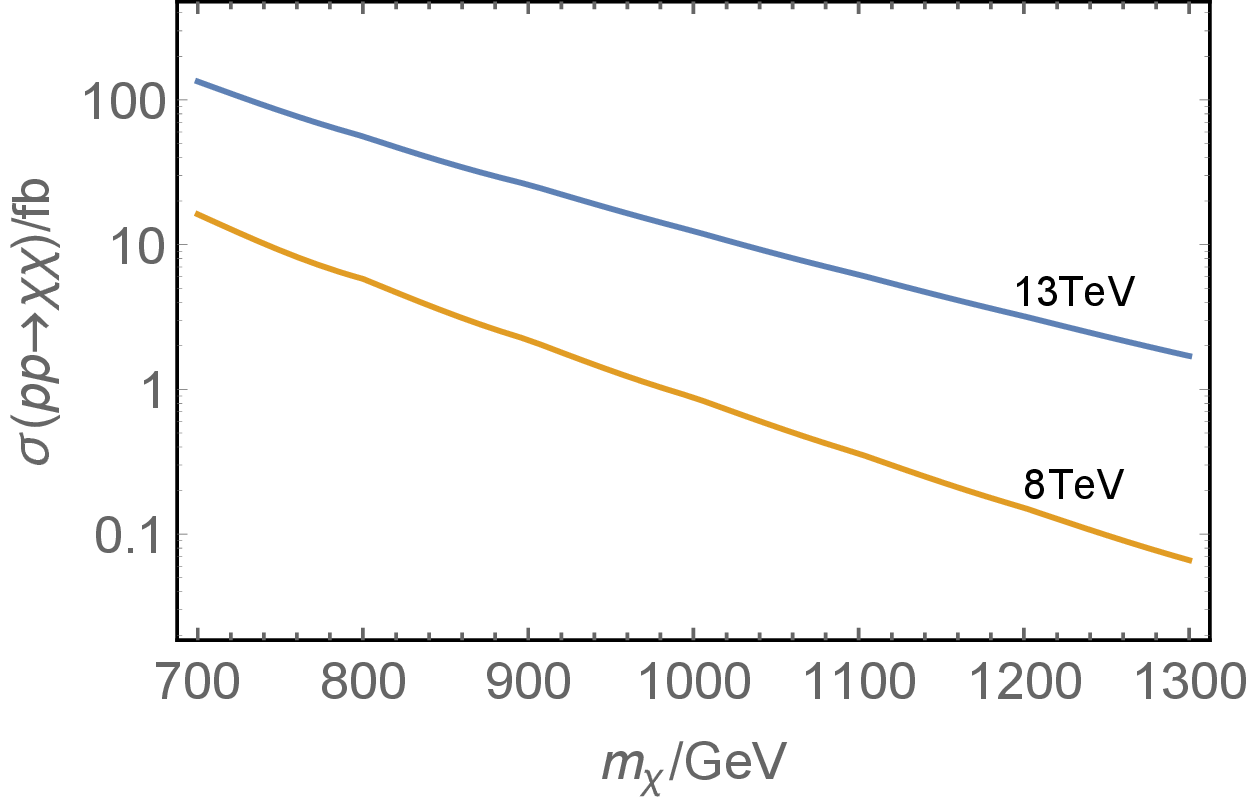}}
\caption{Cross section of single production of $\psi$ through gluon 
 fusion assuming $\sigma(pp \rightarrow \phi \rightarrow \gamma\gamma) 
 = 6~{\rm fb}$ (left) and that of pair production of $\chi$ (right). 
 Both are plotted for $\sqrt{s} = 13~{\rm TeV}$ and $8~{\rm TeV}$.}
\label{fig:psi-chi-prod}
\end{figure}
The $\psi$ particle is created dominantly through single production 
by gluon fusion, whose cross section is plotted in the left panel of 
Fig.~\ref{fig:psi-chi-prod} for $\sigma(pp \rightarrow \phi \rightarrow 
\gamma\gamma) = 6~{\rm fb}$.  (The dependencies on $f$ and $N$ 
cancel for a fixed value of $\sigma(pp \rightarrow \phi \rightarrow 
\gamma\gamma)$.)  Once produced, $\psi$ decays via interactions in 
Eq.~(\ref{eq:pion-couplings}) to $g\gamma$, $gZ$, and $gg$ with the 
branching fractions
\begin{equation}
  B_{\psi \rightarrow g\gamma} 
  = \frac{8g_1^2 \cos^2\!\theta_W}{25g_3^2 + 2g_1^2} 
  \simeq 0.046,
\qquad
  B_{\psi \rightarrow gZ} 
  = \frac{8g_1^2 \sin^2\!\theta_W}{25g_3^2 + 2g_1^2} 
  \simeq 0.014,
\label{eq:psi-decay}
\end{equation}
and $B_{\psi \rightarrow gg} = 1 - B_{\psi \rightarrow gZ} - 
B_{\psi \rightarrow g\gamma}$.  The lower bound on the mass of $\psi$ 
from the LHC data so far~\cite{Aad:2013cva,Aad:2014aqa} is about 
$1~\mbox{--}~1.4~{\rm TeV}$ for $\sigma(pp \rightarrow \phi \rightarrow 
\gamma\gamma) = 4~\mbox{--}~8~{\rm fb}$.  For $N < 5$, this requires 
$r$ to be not significantly smaller than $1$.

The $\chi$ particle is produced only in pairs because of the conservation 
of ``$D$ number'' and ``$L$ number'' at the renormalizable level, 
under which $\Psi_D$ and $\Psi_L$ transform as $(1,0)$ and $(0,1)$, 
respectively.  The production cross section is plotted in the right 
panel of Fig.~\ref{fig:psi-chi-prod}, ignoring possible form factors 
which may become important when the hierarchy between $m_\chi$ and 
$\Lambda$ is not significant due to small $N$, e.g.\ $N = 3$.  The 
signal of $\chi$ depends on its lifetime, which is determined by the 
strength of nonrenormalizable interactions between the hidden quarks 
and the standard model particles violating $D$ and $L$ numbers.  (This 
issue will be discussed in Section~\ref{subsec:chi}.)  Consider first 
the case in which $\chi$ is stable at collider timescales.  In this 
case, a produced $\chi$ particle picks up a light quark of the standard 
model, becoming a heavy fermionic ``hadron.''  Since the charge $\pm 4/3$ 
component of $\chi$ is heavier than the charge $\pm 1/3$ component by 
about $700~{\rm MeV}$~\cite{Cirelli:2005uq}, the former is subject to 
weak decays into the latter with $c\tau \lesssim 1~{\rm mm}$, so that 
the final heavy hadron has a charge $0$ or $\pm 1$.  The mass splitting 
between these neutral and charged hadrons is of order MeV, so that 
the weak decay between them is slow and they can both be regarded 
as stable particles for usual collider purposes.  The lower bound 
on the $\chi$ mass can thus be read off from that of the stable 
bottom squark~\cite{ATLAS:2014fka} with the doubled production 
cross section as given in Fig.~\ref{fig:psi-chi-prod} (because 
of the twice larger number of degrees of freedom).  The bound is 
about $900~{\rm GeV}$.  On the other hand, if the nonrenormalizable 
interactions are strong, $\chi$ may decay promptly into a quark 
and a lepton.  In this case, the lower bound on the $\chi$ mass 
is about $750~{\rm GeV}\mbox{--}1~{\rm TeV}$, depending on the details 
of the decay modes~\cite{Khachatryan:2014ura,Aad:2015caa}.  In any 
event, because of the theoretical expectations in Eq.~(\ref{eq:psi-chi}), 
searches of the $\psi$ and $\chi$ particles provide important probes 
of the model.

Finally, the $\varphi$ particle is standard model color singlet, so 
it can be produced only through electroweak processes or decays of 
heavier resonances.  The decay of $\varphi$ occurs through interactions 
of Eq.~(\ref{eq:pion-couplings}).  The $\varphi^\pm$ decays into $W\gamma$ 
and $WZ$ with the branching fractions of $\cos^2\!\theta_W \simeq 0.77$ 
and $\sin^2\!\theta_W \simeq 0.23$, respectively, while $\varphi^0$ decays 
into $\gamma\gamma$, $\gamma Z$ and $ZZ$ with the branching fractions 
$\sin^2(2\theta_W)/2 \simeq 0.35$, $\cos^2(2\theta_W) \simeq 0.30$, and 
$\sin^2(2\theta_W)/2 \simeq 0.35$, respectively.  The current bounds 
on $\varphi^{\pm,0}$ are weak and do not constrain the model further.

\subsection{Hidden Eta Prime}
\label{subsec:higher}

\begin{figure}[t]
\centering
  \includegraphics[clip,width=.49\textwidth]{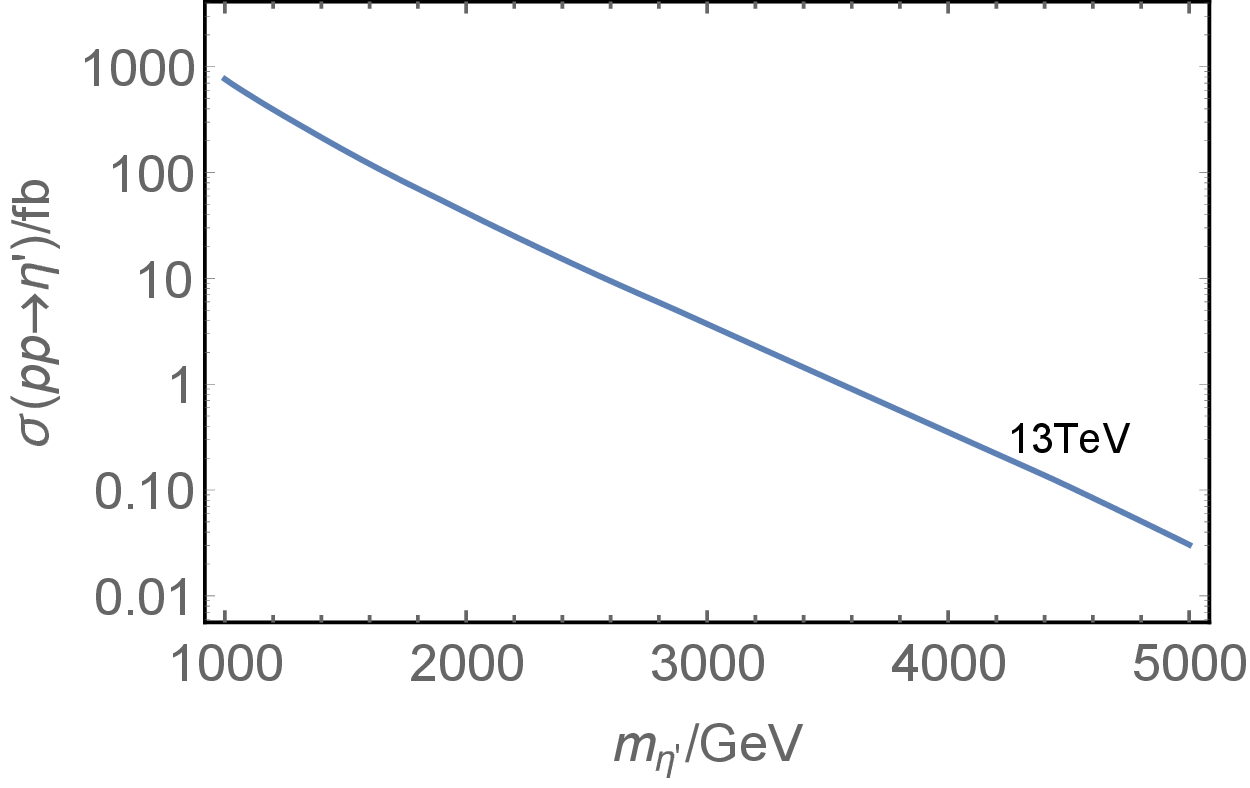}
\caption{Single production cross section of the hidden $\eta'$ 
 through gluon fusion at $\sqrt{s} = 13~{\rm TeV}$, calculated using 
 Eq.~(\ref{eq:eta-couplings}) with $N/f$ determined by Eq.~(\ref{eq:f}) 
 with $\sigma(pp \rightarrow \phi \rightarrow \gamma\gamma) = 
 6~{\rm fb}$.}
\label{fig:eta'-prod}
\end{figure}

Another interesting particle arising from the $G_H$ sector is the hidden 
$\eta'$ state associated with the $U(1)_A$ axial symmetry, which we simply 
call $\eta'$ below.  The mass of this particle is expected to be at the 
dynamical scale~\cite{Witten:1979vv}
\begin{equation}
  m_{\eta'} \approx \sqrt{\frac{5}{N}}\, \Lambda 
  \approx 3.2~{\rm TeV} \sqrt{\frac{6~{\rm fb}}
    {\sigma(pp \rightarrow \phi \rightarrow \gamma\gamma)}},
\label{eq:m_eta'}
\end{equation}
where we have used Eqs.~(\ref{eq:scales},~\ref{eq:f}) in the last 
expression.  The couplings to the standard model gauge bosons can be 
estimated by $U(1)_A$ anomalies with respect to the standard model 
gauge groups
\begin{equation}
  {\cal L} \approx \frac{N g_3^2}{32\sqrt{10}\pi^2 f} \eta'\, 
    \epsilon^{\mu\nu\rho\sigma} G^a_{\mu\nu} G^a_{\rho\sigma} 
  + \frac{N g_2^2}{32\sqrt{10}\pi^2 f} \eta'\, 
    \epsilon^{\mu\nu\rho\sigma} W^\alpha_{\mu\nu} W^\alpha_{\rho\sigma} 
  + \frac{N g_1^2}{32\sqrt{10}\pi^2 f} \eta'\, 
    \epsilon^{\mu\nu\rho\sigma} B_{\mu\nu} B_{\rho\sigma}.
\label{eq:eta-couplings}
\end{equation}
This expression is valid in the large $N$ limit, and we expect that it 
gives a good approximation even for moderately large $N$.  The production 
cross section through gluon fusion calculated using this expression is 
depicted in Fig.~\ref{fig:eta'-prod}.  The energy dependence of the QCD 
$K$ factor is at most of $O(10\%)$ and hence is neglected.

Dominant decay modes of $\eta'$ depend strongly on whether the $G_H$ sector 
respects $CP$ (or parity) or not.  In general, there is the possibility 
that the $G_H$ sector has significant $CP$ violation due to complex phases 
of the hidden quark masses or the $\theta$ parameter for the $G_H$ gauge 
theory.  This possibility is particularly natural if there is a QCD axion, 
eliminating the effect of $CP$ violation on the QCD $\theta$ parameter 
(but no axion acting on the $G_H$ sector); see the Appendix.  In this 
case, $\eta'$ decays into two hidden pions through $CP$ violating terms 
of the form
\begin{equation}
  {\cal L} \sim \frac{m c}{f^3} \eta'\pi\pi,
\label{eq:eta'-pions}
\end{equation}
where $m$ and $\pi$ collectively denote hidden quark masses and hidden 
pion fields, respectively.  The final state then consists of either 
2~quasi-stable particles, $\chi \chi$, or 4~standard model gauge 
bosons.  As discussed in the Appendix, $CP$ violation of the $G_H$ 
sector may also be observed in the neutron electric dipole moment.

If the $G_H$ sector respects $CP$, e.g.\ as in the case in which the 
axion mechanism also operates in the $G_H$ sector, then $\eta'$ decays 
dominantly into 3$\pi$ or 2~standard model gauge bosons.  If the former 
is kinematically open, it dominates the decay; this will be the case 
if the $\eta'$ mass is indeed given by Eq.~(\ref{eq:m_eta'}) with the 
coefficient close to unity.  On the other hand, if the 3$\pi$ mode 
is kinematically forbidden due to a (somewhat unexpected) suppression 
of the coefficient of Eq.~(\ref{eq:m_eta'}), then the decay is to 
2~standard model gauge bosons.  Since the mixing between $\eta'$ 
and hidden pions is suppressed due to small hidden quark masses, 
the branching ratios of $\eta'$ are determined purely by the $SU(5)$ 
flavor symmetry and given by
\begin{align}
  R_{gg} : R_{WW} : R_{ZZ} : R_{Z\gamma} : R_{\gamma\gamma} 
  &\approx \frac{8g_3^4}{g_2^4} : 2 : 
    \biggl( c_W^2 + \frac{5s_W^4}{3c_W^2} \biggr)^2 : 
    2 \biggl( s_W c_W - \frac{5s_W^3}{3c_W} \biggr)^2 : \frac{64 s_W^4}{9}
\nonumber\\
  &\simeq 0.94 : 0.038 : 0.015 : 0.0017 : 0.0072.
\label{eq:eta'-decay-2}
\end{align}
Here, we have used the abbreviations $R_{AB} = B_{\eta' \rightarrow AB}$, 
$s_W = \sin\theta_W$, and $c_W = \cos\theta_W$.  If these modes dominate, 
the production and decay of $\eta'$ also leads to a diphoton signal.%
\footnote{It is amusing to identify this as the origin of the 
 slight ``excess'' at $\simeq 1.6~{\rm TeV}$ in the ATLAS diphoton 
 data~\cite{ATLAS} (although this requires a deviation from the naive 
 estimate of the $\eta'$ mass, Eq.~(\ref{eq:m_eta'}), by a factor of 
 $2$).  Indeed, for $m_{\eta'} = 1.6~{\rm TeV}$, the diphoton rate is 
 about $\sigma(pp \rightarrow \eta' \rightarrow \gamma\gamma) \simeq 
 0.98~{\rm fb}$ (for $\sigma(pp \rightarrow \phi \rightarrow \gamma\gamma) 
 = 6~{\rm fb}$), so we expect a couple of events in the ATLAS data of 
 $3.2~{\rm fb}^{-1}$, which is consistent with the ``excess.''  In this 
 regard, another interesting possibility is that the $G_H$ sector has an 
 extra hidden quark that is singlet under the standard model gauge group. 
 This is discussed in Section~\ref{sec:6-flavor}. \label{ft:1.6TeV}}

\subsection{Heavy Spin-1 Resonances}
\label{subsec:spin-1}

We finally discuss spin-$1$ resonances in the $G_H$ sector having odd 
$C$ and $P$, which we refer to as hidden $\rho$ mesons.  We denote the 
hidden $\rho$ mesons that have the same flavor $SU(5)$ charges as $\psi$, 
$\chi$, $\varphi$, $\phi$, and $\eta'$ by $\rho_\psi$, $\rho_\chi$, 
$\rho_\varphi$, $\rho_\phi$, and $\omega$, respectively.  These hidden 
$\rho$ mesons are expected to be as heavy as $\Lambda$.  They are 
produced at the LHC and yield interesting signals.  For earlier studies 
of phenomenology of spin-$1$ resonant production of pairs of hidden 
particles, see~\cite{Kilic:2009mi,Kilic:2010et}.

The $\rho_\psi$ particle mixes with the standard model gluon and couples 
to standard model quarks with a coupling constant $\sim \sqrt{N} g_3^2 
/4\pi$. The single production cross section of $\rho_\psi$ from quark 
initial states is of order $1000~\mbox{--}~1~{\rm fb}$ for the $\rho_\psi$ 
mass of $2~\mbox{--}~5~{\rm TeV}$ at the $13~{\rm TeV}$ LHC.  It is also 
singly produced from a two gluon initial state via higher dimensional 
operators suppressed by $\Lambda$
\begin{equation}
  {\cal L} \approx i\frac{\sqrt{N}g_3^2}{4\pi\Lambda^2} f^{abc} 
    \left(D_\mu \rho^a_\nu - D_\nu \rho^a_\mu \right) 
    G^b_\rho{}^\mu G^{c\rho \nu}.
\label{eq:L_rho-prod}
\end{equation}
The production cross section from the two gluon initial state is roughly 
comparable to that from quark initial states.  The produced $\rho_\psi$ 
mainly decays into a pair of $\psi$ with a large width, if it is 
kinematically allowed; the resulting $\psi$ in turn decays into $gg$, 
$gZ$, or $g\gamma$, yielding a narrow dijet, $Z$-jet, or $\gamma$-jet 
resonance.  The decay of $\rho_\psi$ into $\psi\phi$ is forbidden 
by $C$-parity.

The $\rho_\chi$ particle is pair produced by ordinary QCD 
interactions.  The pair production cross section is expected to be 
of $O(10^{-1}~\mbox{--}~10^{-9}~{\rm fb})$ for the $\rho_\chi$ mass 
of $2~\mbox{--}~5~{\rm TeV}$ at the $13~{\rm TeV}$ LHC, although there 
may be deviations from this naive QCD estimate by a factor of a few 
due to a form factor.  The produced $\rho_\chi$ decays into $\chi\psi$ 
or $\chi\varphi$, leading to two quasi-stable (or leptoquark-type) 
particles and 4~standard model gauge bosons per event.  The decay of 
$\rho_\chi$ into $\chi\phi$ is forbidden by $C$-parity.

The particles $\rho_\varphi$ and $\rho_\phi$ mix with the standard model 
$SU(2)_L$ and $U(1)_Y$ gauge bosons and couple to standard model quarks 
and leptons with couplings $\sim \sqrt{N} g_2^2 /4\pi$ and $\sim \sqrt{N} 
g_1^2 /4\pi$, respectively.  The single production cross sections of 
$\rho_\varphi$ and $\rho_\phi$ from quark initial states are of order 
$10~\mbox{--}~0.1~{\rm fb}$ and $1~\mbox{--}~0.01~{\rm fb}$, respectively, 
for their masses of $2~\mbox{--}~5~{\rm TeV}$ at the $13~{\rm TeV}$ LHC. 
The coupling between $\rho_\phi$ and two gluons is absent due to $C$-parity. 
The produced $\rho_\varphi$ decays mainly into $\varphi\varphi$, while 
the decay into $\varphi \phi$ is forbidden by $C$-parity.  The $\rho_\phi$ 
decays into $\chi\chi$; the decays into $\psi\psi$, $\varphi\varphi$, 
and $\phi\phi$ are forbidden by $C$-parity.

The $\omega$ particle does not mix with the standard model gauge bosons. 
Furthermore, a coupling between $\omega$ and two gluons is forbidden by 
$C$-parity.  Therefore, the dominant production of $\omega$ occurs through 
a coupling with three gluons
\begin{equation}
  {\cal L} \approx \frac{\sqrt{N} g_3^3}{4\pi \Lambda^4} 
    (\partial_\mu \omega_\nu) d^{abc} G^{a\nu\rho} 
    G^b_{\rho\sigma} G^{c\sigma\mu}.
\label{eq:L_omega-prod}
\end{equation}
In the production process, the initial state is two gluons and the final 
state is a single $\omega$ and a gluon.  The production cross section 
is expected to be of order $10~\mbox{--}~0.01~{\rm fb}$ for the $\omega$ 
mass of $2~\mbox{--}~5~{\rm TeV}$ at the $13~{\rm TeV}$ LHC.  The produced 
$\omega$ decays mainly into three hidden pions and to some extent into 
$\chi\chi$ (with the branching ratio of a few percent, suppressed by 
the size of the flavor $SU(5)$ breaking, $(m_D-m_L)^2/\Lambda^2$); the 
decays into $\psi\psi$, $\varphi\varphi$, and $\phi\phi$ are forbidden 
by $C$-parity and standard model gauge invariance.

\section{Physics at Higher Energies}
\label{sec:higher-e}

Some of the physics of hidden hadrons at the TeV scale are affected 
by theories at higher energies; for example, the lifetime of hidden 
pion $\chi$ is determined by the structure of the theory above $\Lambda$. 
Here we discuss particles that do not decay through $G_H$ or standard 
model gauge dynamics, in particular $\chi$ and low-lying hidden baryons. 
We discuss low-energy operators necessarily to make these particles 
decay and study their phenomenological implications, including constraints 
from cosmology.  We also discuss possible ultraviolet structures that 
lead to the required size of the coefficients of these operators.

\subsection{Physics of Hidden Pion {\boldmath $\chi$}}
\label{subsec:chi}

The dynamics of $G_H$ by itself leaves hidden pion $\chi$ absolutely 
stable.  The $\chi$ particle, however, can decay into standard model 
particles through direct interactions between the $G_H$ and standard 
model sectors.  Here we study physics of $\chi$ decays, focusing on issues 
such as bounds from cosmology and proton decay as well as implications 
for collider physics and theories at very high energies.%
\footnote{Physics of a composite pseudo~Nambu-Goldstone boson that 
 has the same standard model gauge quantum numbers as $\chi$ was 
 discussed in~\cite{Nomura:2005qg}.}

At the level of the standard model fermion bilinears and hidden quark 
bilinears, the most general operators relevant for $\chi$ decays are 
given by
\begin{equation}
  {\cal L} \sim 
  \left\{ \begin{array}{l}
     \Psi_L \sigma_\mu \Psi_D^\dagger \\
     \bar{\Psi}_D \sigma_\mu \bar{\Psi}_L^\dagger 
  \end{array} \right\} \times 
  \left\{ \begin{array}{l}
    q \sigma^\mu u^\dagger \\
    e \sigma^\mu q^\dagger \\ 
    d \sigma^\mu l^\dagger 
  \end{array} \right\}
  + {\rm h.c.},
\label{eq:dim6-1}
\end{equation}
where $q(\Box,\Box,1/6)$, $u(\bar{\Box},{\bf 1},-2/3)$, 
$d(\bar{\Box},{\bf 1},1/3)$, $l({\bf 1},\Box,-1/2)$, and 
$e({\bf 1},{\bf 1},1)$ are the standard model (left-handed Weyl) fermions. 
Since the operators in the first bracket are matched on to $\partial_\mu 
\chi$, these operators give
\begin{equation}
  {\cal L} = \alpha_1 (\partial_\mu \chi) (q \sigma^\mu u^\dagger) 
    + \alpha_2 (\partial_\mu \chi) (e \sigma^\mu q^\dagger) 
    + \alpha_3 (\partial_\mu \chi) (d \sigma^\mu l^\dagger) 
    + {\rm h.c.},
\label{eq:dim6-2}
\end{equation}
at the scale $\Lambda$.  Here, $\alpha_{1,2,3}$ are coefficients of order
\begin{equation}
  \alpha_{1,2,3} \sim \frac{\sqrt{N}}{4\pi} \frac{\Lambda}{M_*^2},
\label{eq:alpha}
\end{equation}
where $M_*$ ($\gtrsim \Lambda$) is the scale at which the operators in 
Eq.~(\ref{eq:dim6-1}) are generated.  We note that since $\Psi_L \sigma_\mu 
\Psi_D^\dagger$ and $\bar{\Psi}_D \sigma_\mu \bar{\Psi}_L^\dagger$ in 
Eq.~(\ref{eq:dim6-1}) correspond to conserved currents in the $G_H$ sector, 
coefficients $\alpha_{1,2,3}$ are given by Eq.~(\ref{eq:alpha}) even if 
the $G_H$ gauge theory is strongly coupled between $M_*$ and $\Lambda$. 
With Eqs.~(\ref{eq:dim6-2},~\ref{eq:alpha}), the decay width of $\chi$ 
is given by
\begin{equation}
  \Gamma_\chi \approx \frac{1}{8\pi} \alpha_{1,2,3}^2 m_f^2 m_\chi 
    \sim \frac{N}{(4\pi)^3} \frac{m_f^2 \Lambda^2}{M_*^4} m_\chi,
\label{eq:Gamma_chi}
\end{equation}
where $m_f$ is the larger of the final state standard model fermion 
masses, arising from chirality suppression.

As we will see in Section~\ref{subsec:cosmo}, cosmology 
requires the lifetime of $\chi$ to be smaller than $\approx 
O(10^{13}\mbox{--}10^{15}~{\rm sec})$, assuming the standard thermal 
history below temperature of about a TeV.  If the operators in 
Eq.~(\ref{eq:dim6-2}) are the only ones contributing to $\chi$ 
decays, then this gives
\begin{equation}
  M_* \lesssim 10^{12}\mbox{--}10^{13}~{\rm GeV} \times \sqrt{\frac{N}{5}}\, 
    \biggl( \frac{m_f}{173~{\rm GeV}} \biggr)^{\frac{1}{2}}
    \biggl( \frac{\Lambda}{3.2~{\rm TeV}\sqrt{N/5}} \biggr)^{\frac{1}{2}} 
    \frac{m_\chi}{1~{\rm TeV}},
\label{eq:cosmo-M*-1}
\end{equation}
where we have used the last expression of Eq.~(\ref{eq:Gamma_chi}) 
and normalized $m_f$ by the top quark mass.  On the other hand, loops 
of hidden quarks involving Eq.~(\ref{eq:dim6-1}) induce standard model 
four-fermion operators suppressed by $\sim (4\pi M_*)^2$.  If the 
coefficients of Eq.~(\ref{eq:dim6-1}) corresponding to $\alpha_1$ and 
one of $\alpha_{2,3}$ are nonzero, this induces proton decay which is 
too rapid for the values of $M_*$ satisfying Eq.~(\ref{eq:cosmo-M*-1}). 
This implies that if $\chi$ decays satisfy the cosmological bound 
due to Eq.~(\ref{eq:dim6-2}), then it must be that $\alpha_1 = 0$ 
or $\alpha_2 = \alpha_3 = 0$.  In this case, $M_*$ may be small enough 
that $\chi$ decays within the detector (even promptly).

One might think that the upper bound on $M_*$ in Eq.~(\ref{eq:cosmo-M*-1}) 
implies that unless early universe cosmology is exotic, there must be new 
physics directly connecting the $G_H$ and standard model sectors well below 
the conventional unification scale, $\sim 10^{14}\mbox{--}10^{17}~{\rm GeV}$. 
This is, however, not necessarily true.  If neither $\Psi_{D,L}$ 
nor $\bar{\Psi}_{D,L}$ is unified in a single multiplet at the high 
energy scale (as in the case of higher dimensional grand unified 
theories~\cite{Hall:2001pg}), we can consider operators leading to 
$\chi$ decays involving standard model operators of dimension~4:
\begin{equation}
  {\cal L} \sim 
  \left\{ \begin{array}{l}
     \bar{\Psi}_D \Psi_L \\
     \Psi_D^\dagger \bar{\Psi}_L^\dagger 
  \end{array} \right\} \times 
  \left\{ \begin{array}{ll}
    q q h^\dagger, & u e h^\dagger \\
    d e h, & q^\dagger l^\dagger h^\dagger \\ 
    u^\dagger u^\dagger h, & u^\dagger d^\dagger h^\dagger
  \end{array} \right\}
  + {\rm h.c.},
\label{eq:dim7-1}
\end{equation}
with $O(1)$ coefficients at the unification scale $M_*$.  (If $\Psi_{D,L}$ 
or $\bar{\Psi}_{D,L}$ is unified, operators of this form with $O(1)$ 
coefficients lead to too large mass terms for the hidden quarks at 
the radiative level.)  These operators lead to operators at the scale 
$\Lambda$
\begin{equation}
  {\cal L} = \beta_1 \chi q q h^\dagger + \beta_2 \chi u e h^\dagger 
    + \beta_3 \chi d e h + \beta_4 \chi q^\dagger l^\dagger h^\dagger 
    + \beta_5 \chi u^\dagger u^\dagger h 
    + \beta_6 \chi u^\dagger d^\dagger h^\dagger 
    + {\rm h.c.},
\label{eq:dim7-2}
\end{equation}
where $h({\bf 1},\Box,-1/2)$ is the standard model Higgs field, and 
$\beta_{1,\cdots,6}$ are coefficients.  An important point is that 
since the operators in the first bracket in Eq.~(\ref{eq:dim7-1}) are 
scalar, their dimensions can receive large quantum corrections if the 
$G_H$ theory is strongly coupled below $M_*$.  Suppose that the $G_H$ 
theory is in a conformal phase between $M_*$ and $M_I$ 
($\gtrsim \Lambda$) because of the existence of extra matter of mass 
$M_I$ charged under $G_H$.  In this case, the size of the coefficients 
in Eq.~(\ref{eq:dim7-2}) is estimated as
\begin{equation}
  \beta_{1,\cdots,6} \sim \frac{\sqrt{N}}{4\pi} 
    \frac{\Lambda^2}{M_*^3} \left( \frac{M_*}{M_I} \right)^{3-\Delta},
\label{eq:beta}
\end{equation}
where $\Delta$ ($> 1$) is the operator dimension in the conformal 
phase, which we take to be the same for $\bar{\Psi}_D \Psi_L$ and 
$\Psi_D^\dagger \bar{\Psi}_L^\dagger$ for simplicity.  (They can in 
general be different depending on the structure of the theory in the 
energy interval between $M_*$ and $M_I$.)  It is therefore possible 
that the cosmological bound on $\chi$ decays $\tau_\chi \lesssim 
O(10^{13}\mbox{--}10^{15}~{\rm sec})$, which translates into 
$\beta_{1,\cdots,6}^{-1} \lesssim 10^{21}\mbox{--}10^{22}~{\rm GeV}$, 
is satisfied even if $M_*$ is at the unification scale.  For example, 
if $M_I$ is within an order of magnitude of $\Lambda$, the bound can 
be satisfied for $\Delta \lesssim 1.2~\mbox{--}~1.7$, with the precise 
value depending on $M_*$, $M_I$, and so on.

The large size of the anomalous dimension leading to small $\Delta$ 
requires that $G_H$ is strongly coupled in the conformal phase.  On the 
other hand, if this phase is still the $G_H$ gauge theory, operators 
of the form
\begin{equation}
  (\bar{\Psi} \Psi) (\bar{\Psi} \Psi)^\dagger,
\label{eq:relevant-1}
\end{equation}
must have dimensions larger than or equal to $4$.  Since these dimensions 
are given by $2\Delta$ in the large $N$ limit (at least for the ones 
directly related to $\bar{\Psi}_D \Psi_L$ or $\bar{\Psi}_L \Psi_D$ 
but also for others if the conformal phase respects the $SU(5)$ flavor 
symmetry), an $O(1/N)$ correction must play the role for making $\Delta$ 
smaller than $2$.  This seems to imply that it is difficult to have 
$M_*$ in the large side, $M_* \approx 10^{17}~{\rm GeV}$, which requires 
$\Delta \lesssim 1.2$.  Furthermore, strong coupling is also expected 
to give large anomalous dimensions for operators of the form
\begin{equation}
  (\bar{\Psi} \Psi)^2.
\label{eq:relevant-2}
\end{equation}
The dimensions of these operators must also be larger than $4$ unless 
their coefficients are strongly suppressed at $M_*$, since otherwise the 
theory flows into a different phase above the scale $\Lambda$, so that 
our low energy theory is not the one discussed in Section~\ref{sec:model}. 
Avoiding the cosmological bound on $\chi$ using only operators in 
Eq.~(\ref{eq:dim7-1}) generated at a unification scale requires these 
conditions to be satisfied.

\subsection{Cosmology}
\label{subsec:cosmo}

Assuming that the thermal history of the universe is standard below 
the temperature of about a TeV, the relic abundance of $\chi$ is 
given as follows.  Below the temperature $T \simeq m_{\chi}/30 \simeq 
30~\mbox{--}~40~{\rm GeV}$, annihilation of $\chi$ into gluons becomes 
ineffective and the $\chi$ abundance freezes out.  If there were no 
subsequent annihilation of $\chi$, this would lead to the present-day 
$\chi$ energy density of $\Omega_\chi \approx O(0.01)$.  However, we 
expect a period of further annihilation at $T \simeq \Lambda_{\rm QCD} 
\approx O(100~{\rm MeV})$ when nonperturbative QCD effects become important. 
At $T \simeq \Lambda_{\rm QCD}$, $\chi$ particles hadronize by picking up 
light quarks, which leads to enhanced annihilation of $\chi$.  While the 
details of this late-time annihilation process are not fully understood, 
we may estimate, based on earlier work~\cite{Kang:2006yd}, that the 
resulting $\chi$ energy density is of order
\begin{equation}
  10^{-8} \lesssim \Omega_\chi \lesssim 10^{-6}.
\label{eq:Omega_chi}
\end{equation}
With this estimate, the possibility of absolutely stable $\chi$ is 
excluded, under reasonable assumptions, by heavy isotope searches; see 
Section~V\,C of Ref.~\cite{Nomura:2005qg} (in which a particle called 
a xyon corresponds to our $\chi$ particle here).

\begin{figure}[t]
\centering
  \includegraphics[clip,width=.55\textwidth]{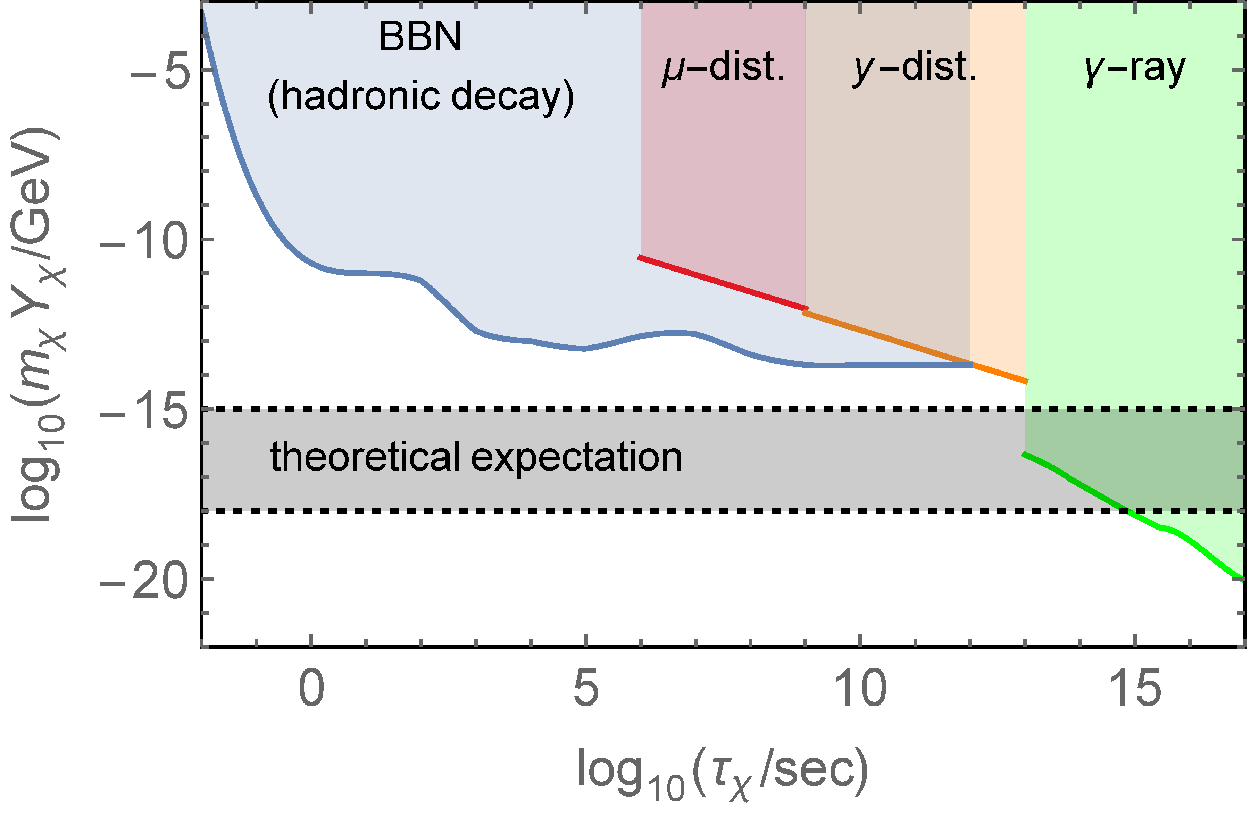}
\caption{Cosmological constraints on the lifetime, $\tau_\chi$, and the 
 product of the mass, $m_\chi$, and entropy yield, $Y_\chi$ of the $\chi$ 
 particle.  The shaded regions are excluded by the analyses of big bang 
 nucleosynthesis (BBN), $\mu$ and $y$ distortions of the cosmic microwave 
 background, and the diffuse gamma-ray background.  The theoretical 
 estimate of $m_\chi Y_\chi$ is indicated by the horizontal band.}
\label{fig:chi-cosmo}
\end{figure}

For unstable $\chi$, bounds on the lifetime are given by the requirement 
that cosmological decay of $\chi$ does neither spoil the success of the 
big bang nucleosynthesis, generate a detectable level of ($\mu$ or $y$) 
distortions of the cosmic microwave background, or lead to an excessive 
amount of the diffuse gamma-ray background.  These constraints, as compiled 
in~\cite{Kawasaki:2007mk} with the update to include~\cite{Ackermann:2014usa}, 
are plotted in Fig.~\ref{fig:chi-cosmo} in the $\tau_\chi$-$m_\chi Y_\chi$ 
plane, where $\tau_\chi$, $m_\chi$, and $Y_\chi = \rho_\chi/s$ are the 
lifetime, mass, and entropy yield of $\chi$.
The estimate of the $\chi$ abundance after the QCD era (corresponding 
to Eq.~(\ref{eq:Omega_chi}) if $\chi$ were absolutely stable) is 
indicated by the horizontal band.  We thus find that with this estimate, 
the bound on the $\chi$ lifetime comes from observations of the diffuse 
gamma-ray background and is given by
\begin{equation}
  \tau_\chi \lesssim 10^{13}\mbox{--}10^{15}~{\rm sec}.
\label{eq:tau_chi-bound}
\end{equation}
Implications of this bound were discussed in Section~\ref{subsec:chi}.

In addition to the $\chi$ particle, the dynamics of $G_H$ leaves 
low-lying hidden baryons stable.  Among all the hidden baryons, we 
expect the lightest ones to be those in which the constituents form 
the smallest spin, i.e.\ $0$ or $1/2$, and have a vanishing orbital 
angular momentum~\cite{Close:1979bt}.%
\footnote{Here we assume that effects involving both orbital and spin 
 angular momenta do not invalidate this expectation, which is indeed 
 the case in QCD.}
These states have masses $\sim N \Lambda$ and lighter than the other 
hidden baryons by $\sim \Lambda$, so all the heavier hidden baryons decay 
into these low-lying hidden baryons by emitting hidden pions and standard 
model gauge bosons.  On the other hand, decays of the low-lying hidden 
baryons require interactions beyond $G_H$ and standard model gauge 
interactions.  In the limit of $m_D = m_L$ and vanishing standard 
model gauge couplings, the low-lying hidden baryons form an $SU(5)$ 
multiplet and are degenerate in mass.  Mass splittings among them, 
therefore, are of order $|m_D - m_L| \approx O(100~{\rm GeV})$ or 
$g_{1,2,3}^2 N \Lambda/16\pi^2 \approx O(10~\mbox{--}~100~{\rm GeV})$, 
so that decays among low-lying hidden baryons cannot occur by emitting 
on-shell $\chi$ particles.  This implies that the dynamics of $G_H$ 
itself leaves all the low-lying hidden baryons stable.

The cosmological fate of the low-lying hidden baryons depends on 
the strength of operators violating hidden baryon number and operators 
responsible for $\chi$ decays.  The former operators lead to decays of 
hidden baryons either directly to standard model particles or to $\chi$ 
and standard model particles through other (off-shell) hidden baryons. 
If the timescale for these decays, $\tau_B$, is shorter than the $\chi$ 
lifetime, $\tau_\chi$, then all the low-lying hidden baryons decay 
in this manner.  On the other hand, if this timescale is longer than 
$\tau_\chi$, the heavier components of the low-lying hidden baryons 
decay first into the lightest hidden baryon (and standard model particles) 
in timescale $\tau_\chi$; then the lightest hidden baryon decays to 
standard model particles (and often $\chi$) in a longer timescale 
of $\tau_B$.

Cosmology of hidden baryons is $N$ dependent, since the spectrum of 
the low-lying hidden baryons as well as operators responsible for their 
decays strongly depend on $N$.  There are, however, some statements one 
can make regardless of the value of $N$.  Suppose $m_D > m_L$.  (The 
other case is discussed later.)  In this case, among the low-lying hidden 
baryons the one composed only of $\Psi_L$ is the lightest.  Then, for 
even and odd $N$ the lightest hidden baryon has standard model gauge 
quantum numbers of $({\bf 1},{\bf 1})_{-N/2}$ and $({\bf 1},\Box)_{-N/2}$, 
respectively, and hence is electrically charged.  The thermal abundance 
of the lightest hidden baryon is determined by its annihilation into 
hidden pions and given by $n_B/s \sim 10^{-16} \times (m_B/10~{\rm TeV})$. 
An electrically charged particle with such an abundance is excluded 
by searches for charged massive stable particles~\cite{DeRujula:1989fe}. 
Thus, the lightest hidden baryon must be unstable in this case.

We now discuss physics of hidden baryon decays in more detail.  For 
illustrative purposes, here we consider the cases of $N=3,4,5$ for 
$m_D > m_L$.  The cases of higher $N$ can be analyzed analogously.

\paragraph*{\boldmath $N=3:$}
The lightest hidden baryon, which is $LLL$ in obvious notation, can decay 
via interactions of the form
\begin{equation}
  {\cal L} \sim 
    u \bar{\Psi}_D \bar{\Psi}_L \bar{\Psi}_L 
    + q \bar{\Psi}_D \bar{\Psi}_D \bar{\Psi}_L 
    + e \bar{\Psi}_D \bar{\Psi}_D \bar{\Psi}_D + {\rm h.c.},
\label{eq:B-1_N=3}
\end{equation}
which induce mixings between the $DLL$, $DDL$, and $DDD$ hidden baryons 
and the right-handed up-type quarks, left-handed quarks, and right-handed 
leptons, respectively:
\begin{equation}
  {\cal L} = \gamma_1 u \bar{B}_{DLL} + \gamma_2 q \bar{B}_{DDL} 
    + \gamma_3 e \bar{B}_{DDD} + {\rm h.c.},
\qquad
  \gamma_{1,2,3} \sim \frac{\sqrt{N}}{(4\pi)^2}
  \frac{\Lambda^3}{M_*^2} \biggl( \frac{M_*}{M_I} \biggr)^{9/2-\Delta_B}.
\label{eq:B-2_N=3}
\end{equation}
Here, we have included possible enhancement of the coefficients by 
conformal dynamics between the scales $M_*$ and $M_I$, with $\Delta_B$ 
($>3/2$) representing the dimension of hidden baryonic operators 
$\bar{\Psi} \bar{\Psi} \bar{\Psi}$ in the conformal phase.%
\footnote{In our analysis here, we assume that the dimensions of different 
 hidden baryonic operators are (approximately) the same.  In particular, 
 we assume that the differences of the dimensions are small enough that 
 operators responsible for hidden baryon decays are given by those in 
 which the sum of the canonical dimensions of the standard model fields 
 is the smallest.}
The lightest hidden baryon $LLL$ then decays into $\chi$ and standard 
model fermions with the decay rate
\begin{equation}
  \Gamma_{LLL} \sim \frac{4\pi \gamma_{1,2,3}^2}{\Lambda}.
\label{eq:B-3_N=3}
\end{equation}
Since this state decays hadronically with the abundance of $\rho_{B}/s 
\sim 10^{-12}~{\rm GeV} \times (m_B/10~{\rm TeV})^2$, preserving 
the success of the big bang nucleosynthesis requires the lifetime 
to be shorter than $O(100~{\rm sec})$~\cite{Kawasaki:2004qu}, which 
translates into $\gamma_{1,2,3} \gtrsim 10^{-12}~{\rm GeV}$.  For 
$M_* \sim 10^{14}\mbox{--}10^{17}~{\rm GeV}$ and $M_I \sim \Lambda$, 
this requires $\Delta_B \lesssim 3.4~\mbox{--}~3.7$.

The decay of the lightest hidden baryon produces $\chi$.  If $\tau_{LLL} 
\equiv \Gamma_{LLL}^{-1} < 10^{-4}~{\rm sec}$, the produced $\chi$ is 
subject to QCD enhanced annihilation afterward, so that the analysis 
in the previous subsection persists.  On the other hand, if $\tau_{LLL} 
> 10^{-4}~{\rm sec}$, the abundance of $\chi$ is determined by its 
annihilation just after the decay of the lightest hidden baryon, which 
is roughly given by
\begin{equation}
  \frac{\rho_\chi}{s} 
  \sim m_\chi \frac{1}{(\Gamma_{LLL} M_{\rm Pl})^{3/2}} 
    \frac{\Gamma_{LLL}}{\sigma_{\chi\bar{\chi}}} 
  \simeq 10^{-13}~{\rm GeV} \frac{m_\chi}{1~{\rm TeV}} 
    \biggl( \frac{\tau_{LLL}}{100~{\rm sec}} \biggr)^{1/2} 
    \frac{1~{\rm fm}^2}{\sigma_{\chi \bar{\chi}}},
\label{eq:B-4_N=3}
\end{equation}
where $M_{\rm Pl}$ is the reduced Planck scale.  The bounds on $\tau_\chi$ 
in this case can be read off from Fig.~\ref{fig:chi-cosmo}.

We note that the hidden baryon number violating operators discussed 
here also induce decays of $\chi$ as they violate $D$ and $L$ numbers. 
Through the mixing between hidden baryons and standard model fermions, 
$\chi$ decays into $u q^\dagger$ or $q e^\dagger$ with the decay rate
\begin{equation}
  \Gamma_\chi \sim 4\pi N m_\chi 
    \Bigl( \frac{\gamma_{1,2,3}}{N\Lambda} \Bigr)^4.
\label{eq:B-5_N=3}
\end{equation}
If $\gamma_{1,2,3} \gtrsim 10^{-8}\mbox{--}10^{-7}~{\rm GeV}$, the lifetime 
of $\chi$ can be shorter than $O(10^{13}\mbox{--}10^{15}~{\rm sec})$ 
without having the operators discussed in the previous section.  For 
$M_* \sim 10^{14}\mbox{--}10^{17}~{\rm GeV}$ and $M_I \sim \Lambda$, 
this translates into $\Delta_B \lesssim 3.1~\mbox{--}~3.3$.  For 
such large $\gamma_{1,2,3}$, hidden baryons are short-lived and do not 
cause any cosmological problems.

\paragraph*{\boldmath $N=4:$}
The lightest hidden baryon, $LLLL$, can decay via
\begin{equation}
  {\cal L} \sim h \Psi_D \Psi_D \Psi_D \Psi_L + {\rm h.c.}
\label{eq:B-1_N=4}
\end{equation}
This induces a mixing between the $DDDL$ hidden baryon and the standard 
model Higgs field
\begin{equation}
  {\cal L}= \delta\, h B_{DDDL} + {\rm h.c.},
\qquad
  \delta \sim \frac{\sqrt{N}}{(4\pi)^3} \frac{\Lambda^5}{M_*^3} 
    \biggl( \frac{M_*}{M_I} \biggr)^{6-\Delta_B},
\label{eq:B-2_N=4}
\end{equation}
where $\Delta_B$ ($> 1$) is the dimension of the operator $\Psi_D \Psi_D 
\Psi_D \Psi_L$ in the conformal phase.  The lightest hidden baryon then 
decays into three $\chi$ and a standard model Higgs or gauge boson, 
with the decay rate
\begin{equation}
  \Gamma_{LLLL} \sim 4\pi \frac{\delta^2}{N^2 \Lambda^3}.
\label{eq:B-3_N=4}
\end{equation}
As the standard model Higgs and gauge bosons decay hadronically, 
the lifetime of the lightest hidden baryon must be shorter than 
$O(100~{\rm sec})$, which requires $\delta \gtrsim 10^{-8}~{\rm GeV}^2$. 
For $M_* \sim 10^{14}\mbox{--}10^{17}~{\rm GeV}$ and $M_I \sim \Lambda$, 
this translates into $\Delta_B \lesssim 3.8~\mbox{--}~4.0$.  Cosmology 
of $\chi$ produced by the lightest hidden baryon decay is as in the 
case of $N=3$.

\paragraph*{\boldmath $N=5:$}
The lightest hidden baryon, $LLLLL$, can decay via interactions of 
the form
\begin{equation}
  {\cal L} \sim h^\dagger d\, \Psi_D \Psi_D \Psi_L \Psi_L \Psi_L 
    + h^\dagger l\, \Psi_D \Psi_D \Psi_D \Psi_L \Psi_L + {\rm h.c.}
\label{eq:B-1_N=5}
\end{equation}
These operators induce couplings between the $DDLLL$ and $DDDLL$ hidden 
baryons with standard model particles
\begin{equation}
  {\cal L} = \epsilon_1 h^\dagger d B_{DDLLL} 
    + \epsilon_2 h^\dagger l B_{DDDLL},
\qquad
  \epsilon_{1,2} \sim \frac{\sqrt{N}}{(4\pi)^4} \frac{\Lambda^6}{M_*^6} 
    \biggl( \frac{M_*}{M_I} \biggr)^{15/2-\Delta_B},
\label{eq:B-2_N=5}
\end{equation}
where $\Delta_B$ ($>3/2$) is the dimension of hidden baryonic operators 
$\Psi\Psi\Psi\Psi\Psi$ in the conformal phase.  The lightest hidden baryon 
decays into a standard model Higgs or gauge boson, a down quark or lepton 
doublet, and multiple $\chi$, with the decay rate
\begin{equation}
  \Gamma_{LLLLL} \sim \frac{N}{4\pi} \epsilon_{1,2}^2 \Lambda.
\label{eq:B-3_N=5}
\end{equation}
As the decay is hadronic, the lifetime of the $LLLLL$ hidden baryon 
must be shorter than $O(100~{\rm sec})$, which requires $\epsilon_{1,2} 
\gtrsim 10^{-15}$.  For $M_* \sim 10^{14}\mbox{--}10^{17}~{\rm GeV}$ 
and $M_I \sim \Lambda$, this translates into $\Delta_B \lesssim 
2.3~\mbox{--}~2.5$.  Cosmology of $\chi$ produced by the decay can 
be analyzed as in the case of $N=3$.

We now discuss the case with $m_D < m_L$.  In this case, depending 
on $N$ and the precise values of $m_{D,L}$, the lightest hidden baryon 
may carry standard model color.  Since a colored hidden baryon is subject 
to late-time annihilation around the QCD phase transition era, the bound on 
its lifetime is weak, $\tau \lesssim O(10^{13}\mbox{--}10^{15}~{\rm sec})$. 
This may allow for the coefficients of the hidden baryon number violating 
operators to be much smaller than the case discussed above.  If $\chi$ 
decay is prompt, this is indeed the case because then all the heavier 
low-lying hidden baryons decay into the lightest hidden baryon before 
the QCD annihilation era.  On the other hand, if $\chi$ is long-lived, 
decays of the low-lying hidden baryons are all controlled by the hidden 
baryon number violating operators.  The bounds on the coefficients are 
then (essentially) the same as before, since they are determined by 
the decays of non-colored hidden baryons.

As seen here, cosmology of hidden baryons is controlled by the lowest 
dimensional hidden baryon number violating operator.  Since its dimension 
depends on the whole hidden quark content, the existence of an extra 
hidden quark could alter the situation.  In particular, if there is 
an extra hidden quark charged under $G_H$ but singlet under $G_{\rm SM}$, 
then the decays of hidden baryons can be faster.  This case will be 
discussed in Section~\ref{sec:6-flavor}.

\subsection{Possible Ultraviolet Theories}
\label{subsec:UV}

We have discussed physics of ``would-be'' stable particles:\ $\chi$ and 
low-lying hidden baryons.  Assuming the standard thermal history of the 
universe below the TeV scale, we have found that these particles must 
decay fast enough via non-renormalizable interactions.  The required 
strength of these interactions suggests the existence of ultraviolet 
physics beyond the minimal $G_H$ and standard model sectors not too 
far above the dynamical scale, $\Lambda$.%
\footnote{An alternative possibility is that the thermal history of 
 the early universe is non-standard.  For detailed analyses of the 
 relic abundance of quasi-stable particles in the case that the 
 reheating temperature is very low, see e.g.~\cite{Harigaya:2014waa}.}

One possibility is that physics responsible for the $\chi$ and hidden 
baryon decays is indeed at a scale $M_*$ which is within a few orders 
of magnitudes of $\Lambda$.  Here we consider an alternative possibility 
that physics leading to these decays is at very high energies, e.g.\ at 
a unification scale $M_* \approx O(10^{14}\mbox{--}10^{16}~{\rm GeV})$. 
Here, we consider a few examples that this can be the case.
\begin{itemize}
\item
{\bf Conformal dynamics}

As we have discussed before, even if the relevant non-renormalizable 
interactions are generated far above $\Lambda$, such as at the unification 
scale, conformal dynamics of $G_H$ may enhance these operators and induce 
sufficiently fast decays of the would-be stable particles.  Having conformal 
dynamics requires a sufficient number of (vectorlike) particles charged 
under $G_H$ in addition to $\Psi_{D,L}$ and $\bar{\Psi}_{D,L}$, which 
we may assume to have masses larger than the dynamical scale by a factor 
of $O(1~\mbox{--}~10)$.  In this case, we can consider that $G_H$ is in 
a conformal phase above the mass scale of these particles but deviates 
from it below this mass scale and finally confines at $\Lambda$.  With 
such dynamics, we can understand the proximity of the dynamical scale 
of $G_H$, $\Lambda \sim {\rm TeV}$, and the masses of hidden quarks 
$m \sim 0.1~{\rm TeV}$, if the masses $\Psi_{D,L}$, $\bar{\Psi}_{D,L}$ 
and additional particles originate from a common source and the conformal 
phase of $G_H$ is strongly coupled.  The existence of a particle charged 
under $G_H$ and singlet under $G_{\rm SM}$ may also help satisfying the 
constraints from cosmology; see Section~\ref{sec:6-flavor}.  We note 
that (some of) the additional particles added here may be charged under 
$G_{\rm SM}$, which does not destroy gauge coupling unification if they 
form complete $SU(5)$ multiplets.

\item
{\bf Supersymmetry}

If superpartners of the standard model particles and hidden quarks are 
near the TeV scale, then the decay rates of would-be stable particles 
can be larger than those in the non-supersymmetric model.  (With the 
superpartners near the TeV scale, $N \leq 4$ is required in order for 
the standard model gauge couplings not to blow up below the unification 
scale.)  Specifically, hidden baryon number can be more easily broken. 
For $N=3$, for example, there are dimension-five superpotential operators
\begin{equation}
  W \sim \frac{1}{M_*} \left( u \bar{\Psi}_D \bar{\Psi}_L \bar{\Psi}_L 
    + q \bar{\Psi}_D \bar{\Psi}_D \bar{\Psi}_L 
    + e \bar{\Psi}_D \bar{\Psi}_D \bar{\Psi}_D \right),
\label{eq:SUSY}
\end{equation}
which mix standard model particles with hidden baryons.  Even if $M_*$ 
is around the unification scale, hidden baryons decay with lifetimes 
shorter than $O(100~{\rm sec})$ if superpartner masses are around 
$O(1~\mbox{--}~10~{\rm TeV})$.  Introduction of superparticles at the 
TeV scale also improves gauge coupling unification.

\item
{\bf Superconformal dynamics}

It is possible that both conformal dynamics and supersymmetry are 
at play above some scale not far from $\Lambda$.  This scenario is very 
interesting.  Conformal dynamics as well as exchange of superpartners may 
enhance the decay rates of would-be stable particles, and the deviation 
of $G_H$ from the conformal window may be explained by the decoupling 
of hidden squarks and hidden gauginos at $O(1~\mbox{--}~10~{\rm TeV})$. 
Moreover, the anomalous dimensions of some of the operators can be 
calculated due to supersymmetry.

The ultraviolet structure of this class of theories is tightly 
constrained if the $G_H$ theory is already in a conformal phase 
at the unification scale $M_*$.  Let us first assume that all the 
superpartners are at $\tilde{m} \approx O({\rm TeV})$.  In this case, 
$N \geq 4$ leads to a Landau pole for the standard model gauge couplings 
below $M_*$.  This is because in the energy interval between $\tilde{m}$ 
and $M_*$, the $G_H$ gauge theory is on a nontrivial fixed point, so 
that the effect of $G_H$ gauge interactions enhances the contribution 
of the hidden quarks to the running of the standard model gauge 
couplings (at the level of two loops in the standard model gauge 
couplings).  Even for $N=3$, the number of flavors of the $G_H$ 
gauge theory is bounded from below, since the $G_H$ gauge coupling 
at the fixed point is larger for a smaller number of flavors, making 
the contribution from the hidden quarks larger.  We find that the 
number of flavors must be $7$ or $8$ so that $G_H$ is in the conformal 
window while at the same time the standard model gauge couplings do 
not hit a Landau pole below the unification scale.  Assuming that 
additional particles always appear in the form of complete $SU(5)$ 
multiplets, this predicts the existence of particles that are charged 
under $G_H$ but singlet under $G_{\rm SM}$.  As we will see in 
Section~\ref{sec:6-flavor}, the existence of such particles helps 
to evade constraints from cosmology.  For heavier superparticle 
masses, the possible choice of $N$ and the number of flavors, $F$, 
increases; for example, for $\tilde{m} \approx O(10~{\rm TeV})$, 
$(N,F) = (3,6)$, $(4,10)$, and $(4,11)$ are also allowed.
\end{itemize}

\section{An Extra Hidden Light Quark}
\label{sec:6-flavor}

In this section, we discuss a singlet extension of the model described 
in Section~\ref{sec:model}:\ we add an extra vectorlike hidden quark, 
$\Psi_N$ and $\bar{\Psi}_N$, which is in the fundamental representation 
of $G_H$ but is singlet under the standard model gauge group.  As mentioned 
in Section~\ref{subsec:cosmo}, this makes it easier to avoid constraints 
from cosmology.  It also leads to two diphoton resonances in the spectrum 
of hidden pions, which has interesting phenomenological implications 
(one of which was mentioned in footnote~\ref{ft:1.6TeV}).  The full matter 
content of the model is summarized in Table~\ref{tab:singlet}.
\begin{table}[t]
\begin{center}
\begin{tabular}{c|cccc}
   & $G_H = SU(N)$ & $SU(3)_C$ & $SU(2)_L$ & $U(1)_Y$ \\ \hline
 $\Psi_D$       &       $\Box$ & $\bar{\Box}$ & ${\bf 1}$ &  $1/3$ \\
 $\Psi_L$       &       $\Box$ &    ${\bf 1}$ &    $\Box$ & $-1/2$ \\
 $\Psi_N$       &       $\Box$ &    ${\bf 1}$ & ${\bf 1}$ &    $0$ \\
 $\bar{\Psi}_D$ & $\bar{\Box}$ &       $\Box$ & ${\bf 1}$ & $-1/3$ \\
 $\bar{\Psi}_L$ & $\bar{\Box}$ &    ${\bf 1}$ &    $\Box$ &  $1/2$ \\
 $\bar{\Psi}_N$ & $\bar{\Box}$ &    ${\bf 1}$ & ${\bf 1}$ &    $0$ 
\end{tabular}
\end{center}
\caption{Charge assignment of the extended model. $\Psi_{D,L,N}$ and 
 $\bar{\Psi}_{D,L,N}$ are left-handed Weyl spinors.}
\label{tab:singlet}
\end{table}
This charge assignment was also considered in~\cite{Kilic:2009mi}.  We here 
take the masses of the hidden quarks
\begin{equation}
  {\cal L} = -m_D \Psi_D \bar{\Psi}_D - m_L \Psi_L \bar{\Psi}_L 
    -m_N \Psi_N \bar{\Psi}_N + {\rm h.c.},
\label{eq:SU6-L_mass}
\end{equation}
to be real and positive without loss of generality.  We assume $m_{D,L,N} 
\lesssim \Lambda$.

The spectrum below $\Lambda$ consists of hidden pions
\begin{gather}
  \psi({\bf Adj}, {\bf 1}, 0), \qquad
  \chi\Bigl(\Box, \Box, -\frac{5}{6}\Bigr), \qquad
  \varphi({\bf 1}, {\bf Adj}, 0), \qquad
  \phi({\bf 1}, {\bf 1}, 0),
\nonumber \\
  \xi\Bigl(\Box,{\bf 1},-\frac{1}{3}\Bigr), \qquad
  \lambda\Bigl({\bf 1},\Box,\frac{1}{2}\Bigr), \qquad
  \eta({\bf 1}, {\bf 1}, 0),
\end{gather}
where $\chi$, $\xi$, and $\lambda$ are complex while the others are real. 
Note that there are two hidden pions which are singlet under the standard 
model gauge group, $G_{\rm SM}$---one is charged under $SU(5) \supset 
G_{\rm SM}$ while the other is singlet, which we refer to as $\phi$ and 
$\eta$, respectively.  The masses of the hidden pions are given by
\begin{align}
  m_\psi^2 &= 2 m_D \frac{c}{f^2} 
    + 3 \Delta_C,
\label{eq:SU6-psi}\\
  m_\chi^2 &= (m_D + m_L) \frac{c}{f^2} 
    + \frac{4}{3} \Delta_C + \frac{3}{4} \Delta_L + \frac{5}{12} \Delta_Y,
\label{eq:SU6-chi}\\
  m_\varphi^2 &= 2 m_L \frac{c}{f^2} 
    + 2 \Delta_L,
\label{eq:SU6-varphi}\\
  m_\xi^2 &= (m_D + m_N) \frac{c}{f^2} 
    + \frac{4}{3} \Delta_C+ \frac{1}{15} \Delta_Y,
\label{eq:SU6-xi}\\
  m_\lambda^2 &= (m_L + m_N) \frac{c}{f^2} 
    + \frac{3}{4} \Delta_L+ \frac{3}{20} \Delta_Y,
\label{eq:SU6-lambda}\\
  \left( \begin{array}{cc}
    m_\phi^2 & m_{\phi\eta}^2 \\
    m_{\eta\phi}^2 & m_\eta^2 
  \end{array} \right) 
  &= \left( \begin{array}{cc}
    \frac{2}{5}(2 m_D + 3 m_L) & \frac{2}{5}(m_D - m_L) \\
    \frac{2}{5}(m_D - m_L) & \frac{1}{15}(3 m_D + 2 m_L + 25 m_N) 
  \end{array} \right) \frac{c}{f^2},
\label{eq:SU6-phi-eta}
\end{align}
where $c$ and $f$ are the hidden quark bilinear condensates and the decay 
constant, respectively.

The mixing between $\phi$ and $\eta$ vanishes for $m_D=m_L$ due to the 
$SU(5)$ symmetry.  Expect for this special case, the mass eigenstates 
$\phi_+$ and $\phi_-$ are determined by Eq.~(\ref{eq:SU6-phi-eta})
\begin{equation}
  \phi_+ = \eta \cos\theta + \phi \sin\theta,
\qquad
  \phi_- = -\eta \sin\theta + \phi \cos\theta.
\label{eq:singlet_mix}
\end{equation}
Here, the mixing angle $\theta$ and the mass eigenvalues $m_+$ and $m_-$ 
($m_+^2 > m_-^2$) are related with $m_{D,L,N}$ as
\begin{align}
  m_D &= \frac{f^2}{c} \frac{m_-^2 - 3 (m_+^2-m_-^2)\tan\theta 
    + m_+^2 \tan^2\!\theta}{2 (1 + \tan^2\!\theta)},
\label{eq:SU6-mD}\\
  m_L &= \frac{f^2}{c} \frac{m_-^2 + 2 (m_+^2-m_-^2)\tan\theta 
    + m_+^2 \tan^2\!\theta}{2 (1 + \tan^2\!\theta)},
\label{eq:SU6-mL}\\
  m_N &= \frac{f^2}{c} \frac{6 m_+^2 - m_-^2 + (m_+^2-m_-^2)\tan\theta 
    + (6 m_-^2-m_+^2) \tan^2\!\theta}{10 (1 + \tan^2\!\theta)}. 
\end{align}
The dimension-five couplings of the hidden pions with the standard model 
gauge fields are determined by chiral anomalies.  The couplings of 
$\psi$, $\varphi$, and $\phi$ are given by Eq.~(\ref{eq:pion-couplings}), 
while those of $\eta$ are given by Eq.~(\ref{eq:eta-couplings}) with 
the replacement $\eta' \rightarrow \eta/\sqrt{6}$.  The couplings of 
the mass eigenstates $\phi_{\pm}$ can be read off from these expressions 
and the mixing in Eq.~(\ref{eq:singlet_mix}).

\subsection{Diphoton (Diboson) Signals and Other Phenomenology}
\label{subsec:SU6-diboson}

A distinct feature of this model is that there are two standard model 
singlet hidden pions, $\phi_+$ and $\phi_-$, which are produced via gluon 
fusion and decay into a pair of standard model gauge bosons, including 
a diphoton.  If kinematically allowed, they may also decay into three 
hidden pions; with parity violation by $\theta_H \neq 0$, they also 
decay into two hidden pions.  Here we assume that the decay channels 
into hidden pions are suppressed kinematically and/or by $\theta_H 
\simeq 0$.  There are several interesting possibilities to consider 
in terms of the phenomenology of these particles:
\begin{itemize}
\item
{\bf 1.6~TeV diphoton excess}

We may identify the two singlets $\phi_-$ and $\phi_+$ as the origins 
of, respectively, the $750~{\rm GeV}$ excess and the slight ``excess'' 
at $\simeq 1.6~{\rm TeV}$ seen in the ATLAS diphoton data~\cite{ATLAS}. 
\begin{figure}[t]
\centering
  \subfigure{\includegraphics[clip,width=.50\textwidth]{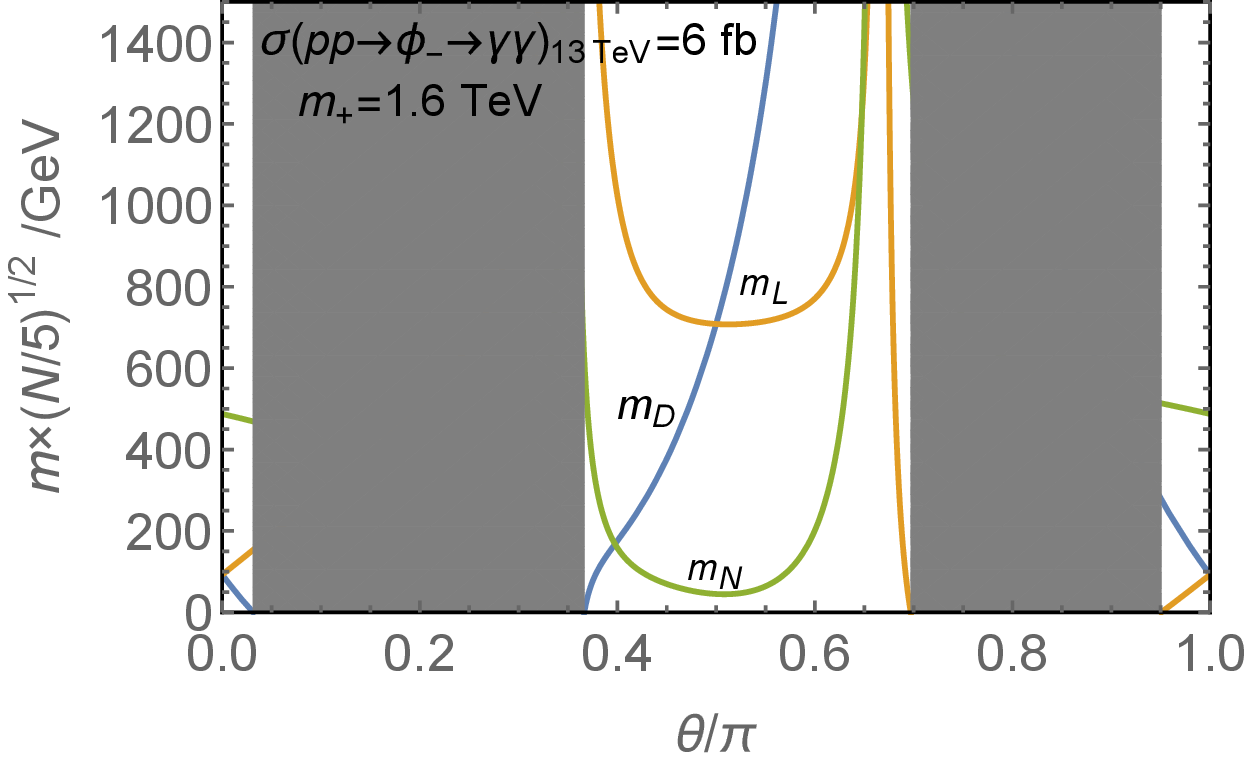}}
\hspace{3mm}
  \subfigure{\includegraphics[clip,width=.47\textwidth]{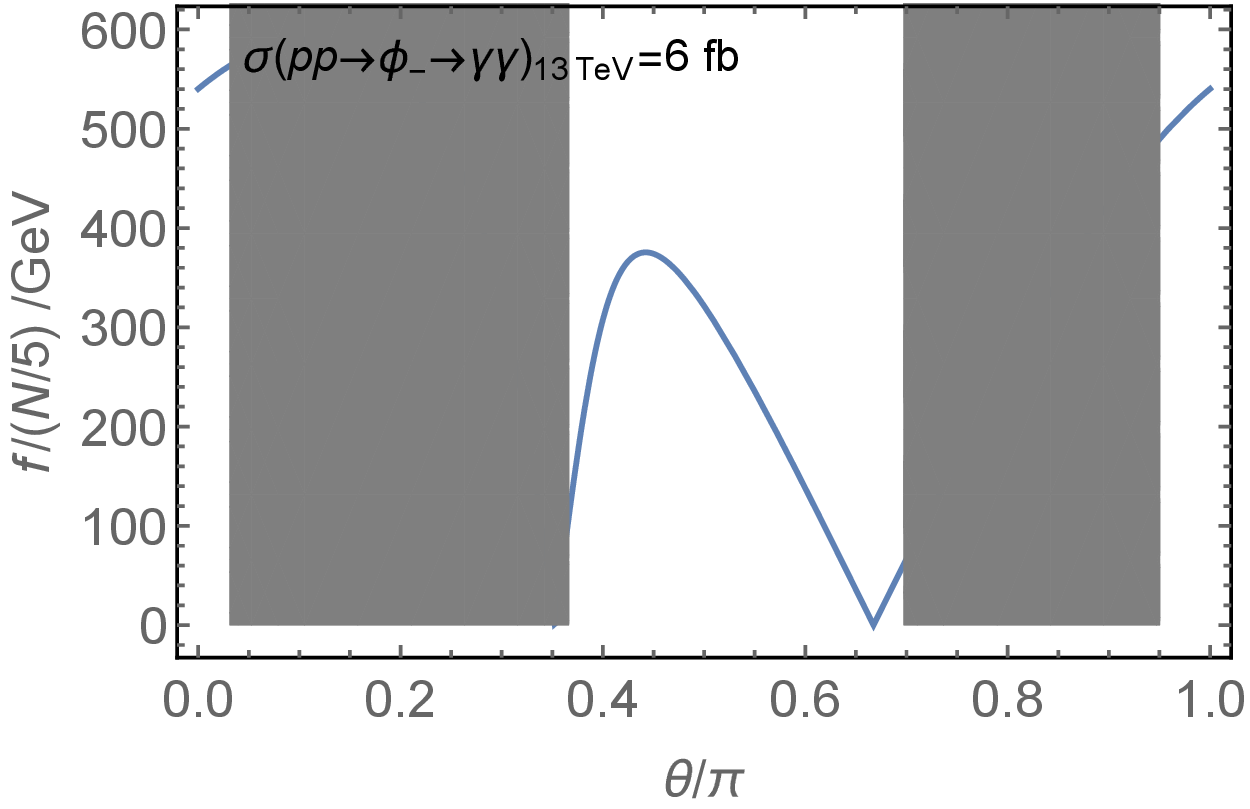}}
\caption{The hidden quark masses, $m_D$, $m_L$ and $m_N$, reproducing 
 $m_- = 750~{\rm GeV}$ and $m_+ = 1.6~{\rm TeV}$ as a function of $\theta$ 
 (left).  Here, the value of the decay constant $f$ is determined so 
 that $\sigma(pp\rightarrow \phi_- \rightarrow \gamma\gamma)=6~{\rm fb}$ 
 is obtained at $\sqrt{s} = 13~{\rm TeV}$ (right).  In the gray-shaded 
 regions of $\theta$, no choice of $m_{D,L,N}$ may reproduce the required 
 $\phi_\pm$ masses.}
\label{fig:theta_m-f}
\end{figure}
In the left panel of Fig.~\ref{fig:theta_m-f}, we show the values 
of the hidden quark masses $m_{D,L,N}$ needed to reproduce $m_- = 
750~{\rm GeV}$ and $m_+ = 1.6~{\rm TeV}$ as a function of $\theta$. 
Here, the decay constant $f$ has been determined so that $\sigma(pp 
\rightarrow \phi_- \rightarrow \gamma\gamma) = 6~{\rm fb}$ at $\sqrt{s} 
= 13~{\rm TeV}$, as plotted in the right panel.  In the parameter 
region $\theta/\pi \in [0,0.03) \cup (0.37,0.70) \cup (0.95, 1)$, 
$m_- = 750~{\rm GeV}$ and $m_+ = 1.6~{\rm TeV}$ can be obtained by 
an appropriate choice of $m_{D,L,N}$, but in the other regions---which 
are shaded---no choice of $m_{D,L,N}$ can lead to the required $\phi_\pm$ 
masses.  Near the edge of the allowed region $(0.37,0.70)$, the decay 
constant $f$ is required to be small.  This is because at $\theta/\pi 
\sim 0.35$ the dimension-five coupling between $\phi_-$ and gluons 
vanishes, while at $\theta/\pi \sim 0.65$ that between $\phi_-$ 
and photons vanishes, so formally $f \rightarrow 0$ is required to obtain 
$\sigma(pp \rightarrow \phi_- \rightarrow \gamma\gamma) = 6~{\rm fb}$ 
at these values of $\theta$.  Note that for $m_{D,L,N} \gtrsim \Lambda 
\simeq 4\pi f/\sqrt{N}$, the results obtained here using chiral perturbation 
theory expressions, Eq.~(\ref{eq:SU6-psi}~--~\ref{eq:SU6-phi-eta}), 
cannot be fully trusted, although we may still regard them as giving 
qualitatively correct estimates.

\begin{figure}[t]
\centering
  \subfigure{\includegraphics[clip,width=.48\textwidth]{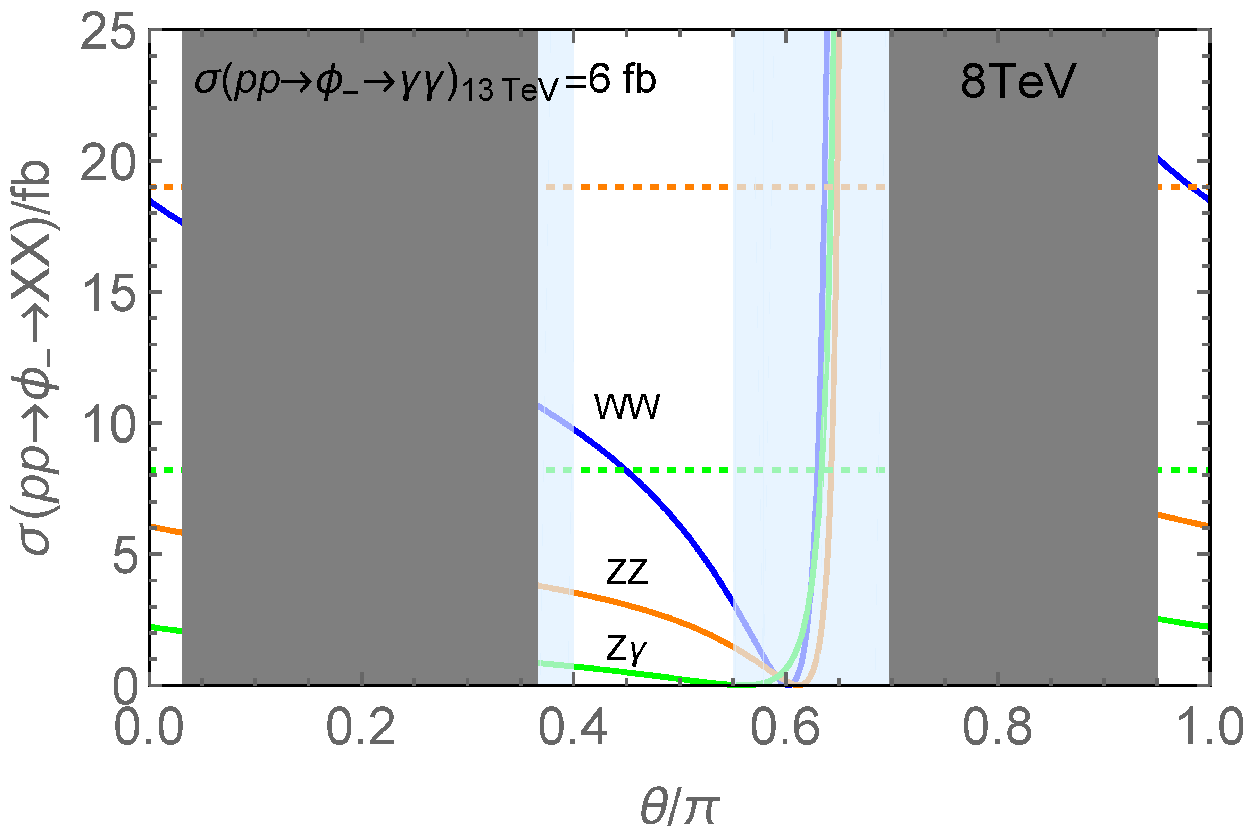}}
\hspace{3mm}
  \subfigure{\includegraphics[clip,width=.49\textwidth]{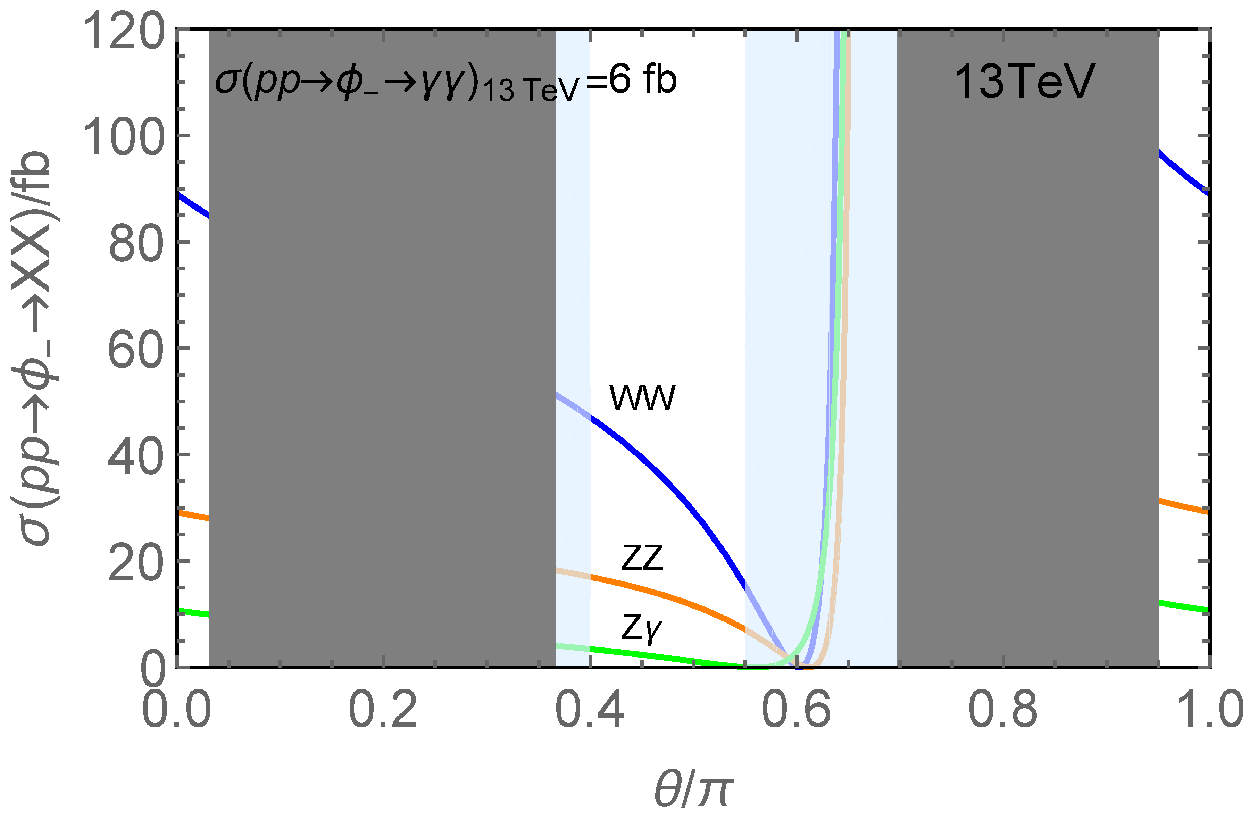}}
\caption{The production cross section of $\phi_-$ times branching ratios 
 into two electroweak gauge bosons at $\sqrt{s} = 8~{\rm TeV}$ (left) 
 and $13~{\rm TeV}$ (right).  The gray-shaded regions are unphysical. 
 In the blue-shaded regions, chiral perturbation theory cannot be fully 
 trusted, and the plots represent only qualitative estimations.}
\label{fig:phi-}
\end{figure}
In Fig.~\ref{fig:phi-}, we present predictions for the production cross 
section of $\phi_-$ times the branching ratios into two electroweak gauge 
bosons at $\sqrt{s} = 8~{\rm TeV}$ (left) and $13~{\rm TeV}$ (right). 
In the blue-shaded regions, chiral perturbation theory gives only 
qualitative estimates because of $m_{D,L,N} \gtrsim \Lambda$.  In the 
left panel, we also show the upper bound on the cross section for each 
mode~\cite{Harigaya:2015ezk,Knapen:2015dap} by the dotted line using 
the same color as the corresponding prediction.  For any $\theta/\pi\in 
[0,0.03) \cup (0.4,0.55) \cup (0.95,1)$, in which chiral perturbation 
theory can be trusted, the bounds are all satisfied.  Predictions for 
the production cross section of the other neutral hidden pion, $\phi_+$, 
times branching ratios into two electroweak gauge bosons at $\sqrt{s} 
= 13~{\rm TeV}$ are presented in Fig.~\ref{fig:phi+_1600}. 
\begin{figure}[t]
\centering
  \includegraphics[clip,width=.49\textwidth]{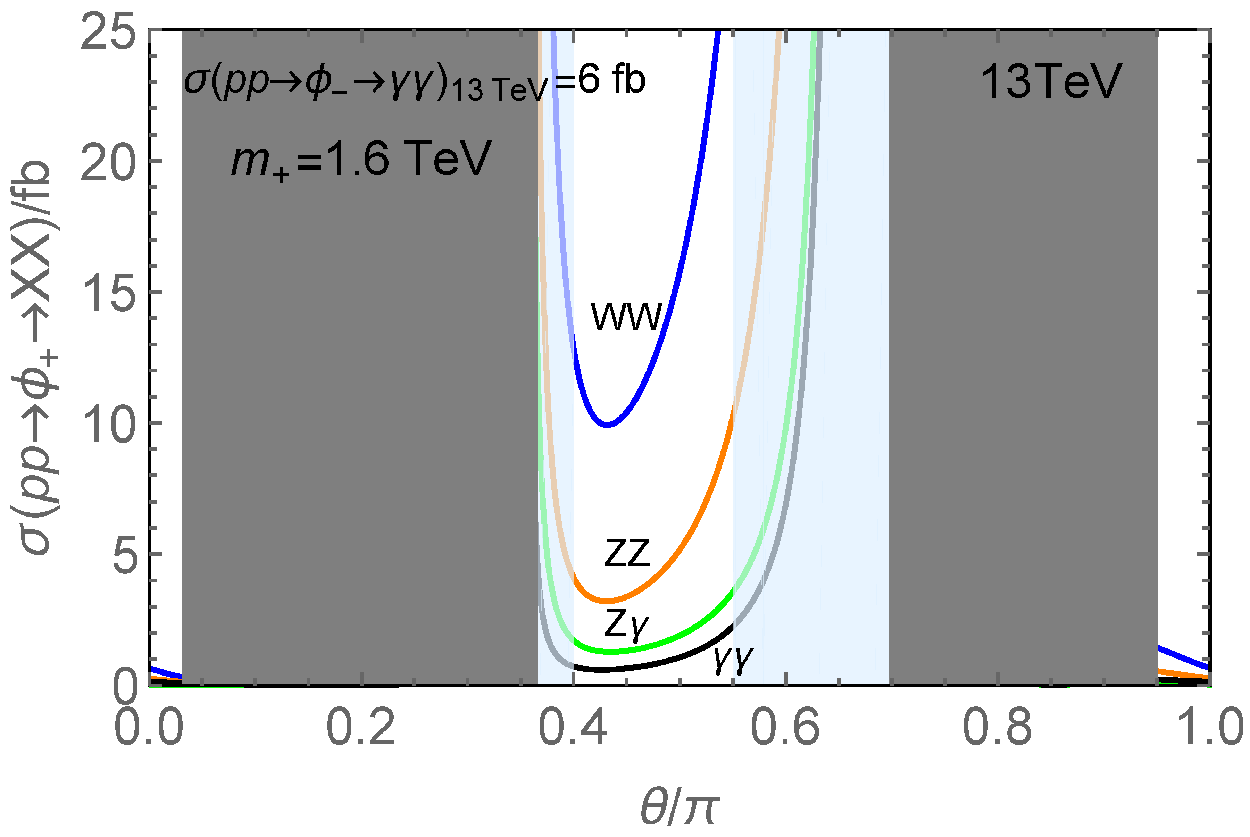}
\caption{The production cross section of $\phi_+$ of mass $m_+ = 
 1.6~{\rm TeV}$ times the branching ratios into two electroweak gauge 
 bosons at $\sqrt{s} = 13~{\rm TeV}$.}
\label{fig:phi+_1600}
\end{figure}
For $\theta \sim \pi/2$, the production cross section of diphoton 
is $O({\rm fb})$, which is consistent with the ``excess'' at $\simeq 
1.6~{\rm TeV}$.  We also show the prediction for the masses of the 
hidden pions in Fig.~\ref{fig:theta-pion}. 
\begin{figure}[t]
\centering
  \includegraphics[clip,width=.50\textwidth]{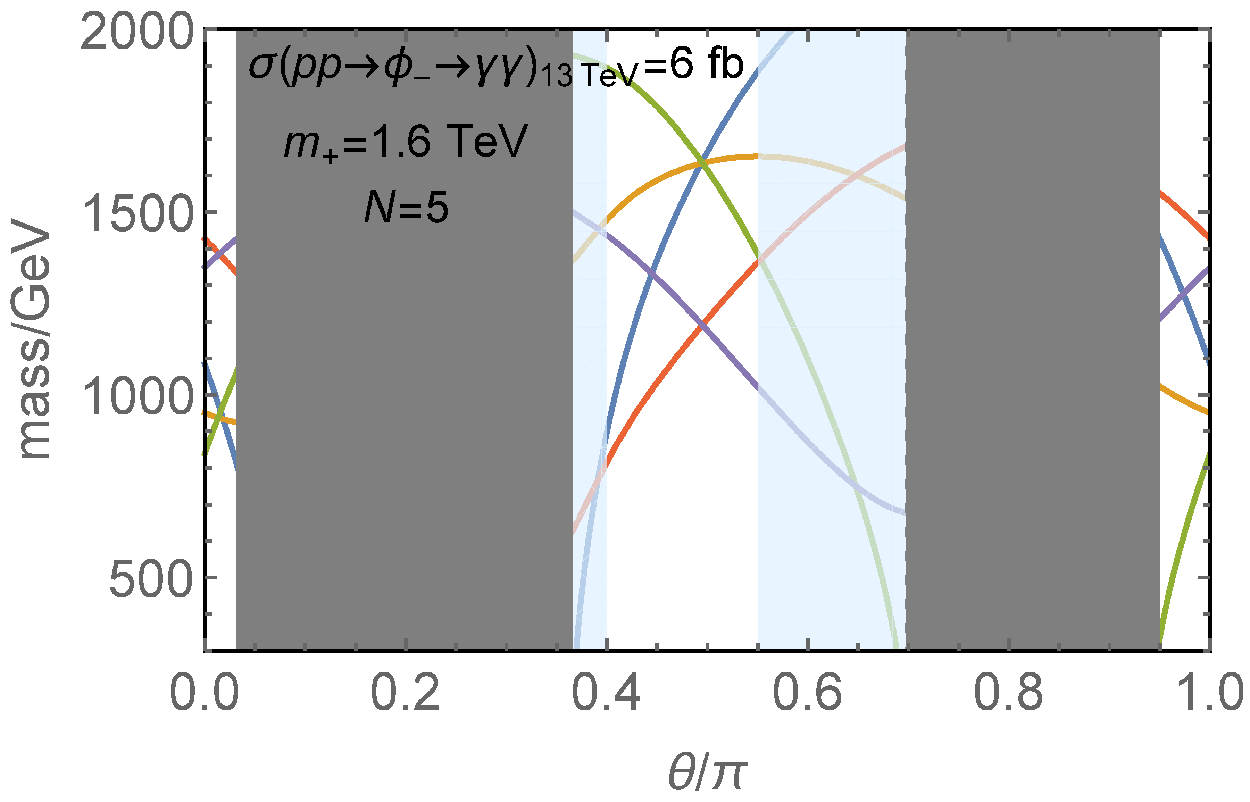}
\caption{The masses of hidden pions $\psi$ (blue), $\chi$ (orange), 
 $\varphi$ (green), $\xi$ (red) and $\lambda$ (purple) for $m_- = 
 750~{\rm GeV}$ and $m_+ = 1.6~{\rm TeV}$ as a function of $\theta$. 
 Here, we have chosen $N=5$.}
\label{fig:theta-pion}
\end{figure}
For $\theta \sim \pi/2$, the mass of the $\chi$ particle can reach 
$\sim 1.5~{\rm TeV}$.  This helps to evade the experimental bound on 
this particle (see Section~\ref{subsec:pions}).

\item
{\bf (Apparent) wide width of the 750~GeV excess}

Alternatively, we may consider that both $\phi_-$ and $\phi_+$ have 
masses around $750~{\rm GeV}$ with a small mass difference of 10s of 
GeV.  In this case, the two resonances are observed as an apparent wide 
resonance~\cite{Franceschini:2015kwy}, which is mildly preferred by the 
ATLAS diphoton data.  Such mass degeneracy occurs if $m_D \simeq m_L 
\simeq m_N$.  The required value of the decay constant $f$ is larger 
than that in Eq.~(\ref{eq:f}):
\begin{equation}
  f \simeq 720~{\rm GeV}\, \frac{N}{5} \sqrt{\frac{6~{\rm fb}} 
    {\sigma(pp \rightarrow \phi \rightarrow \gamma\gamma)}},
\label{eq:f_degenerate}
\end{equation}
because the two resonances contribute to the diphoton rate.  With 
Eqs.~(\ref{eq:SU6-psi}~--~\ref{eq:SU6-phi-eta},~\ref{eq:f_degenerate}) 
and $m_D \simeq m_L \simeq m_N$, the masses of the hidden pions are 
determined for given $N$, which are shown in Fig.~\ref{fig:N-pion}. 
\begin{figure}[t]
\centering
  \includegraphics[clip,width=.48\textwidth]{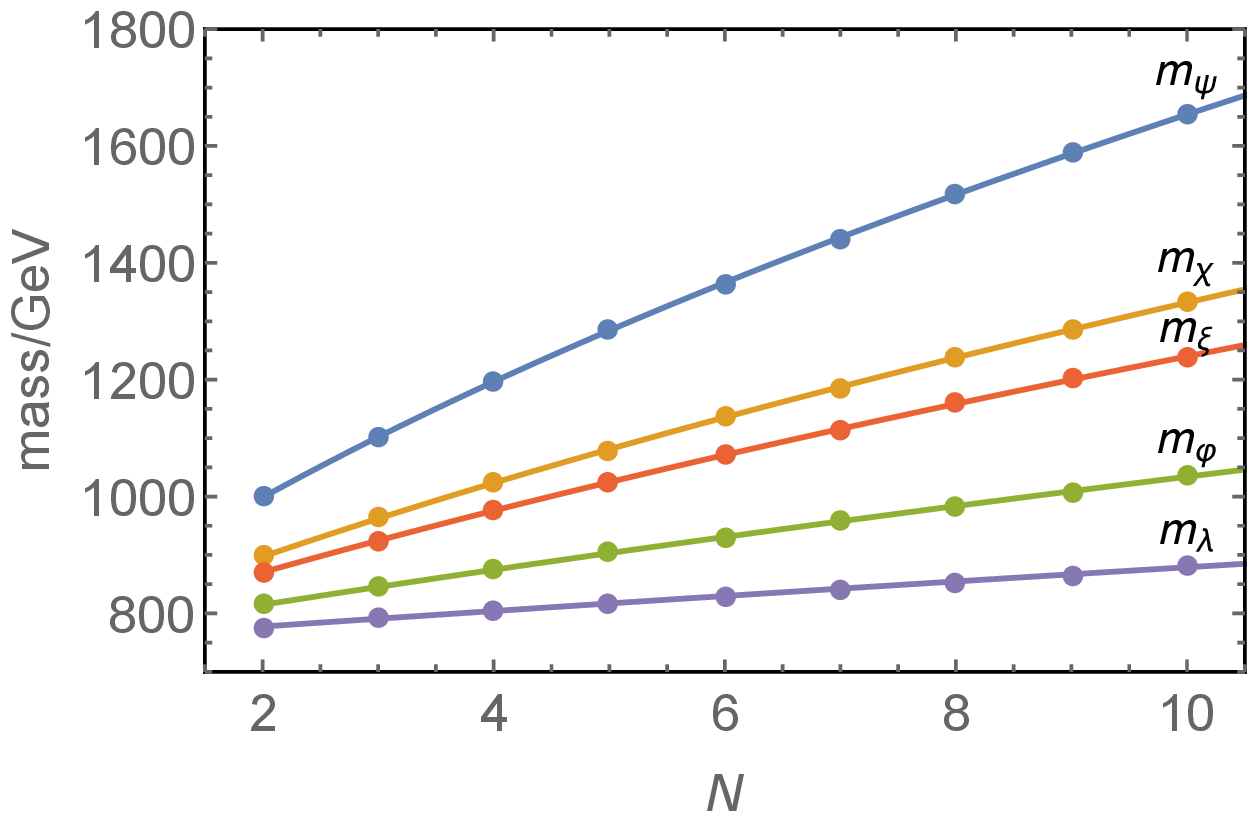}
\caption{The masses of hidden pions $\psi$, $\chi$, $\varphi$, $\xi$, 
 and $\lambda$ for $m_- \simeq m_+ \simeq 750~{\rm GeV}$ for given 
 values of $N$.}
\label{fig:N-pion}
\end{figure}
Since $f$ is larger, the hidden pion masses are larger than in the 
model without the singlet hidden quark.  In particular, $m_\chi$ easily 
exceeds $1~{\rm TeV}$, helping to evade the constraint.

\item
{\bf 2~TeV diboson excess}

Yet another possibility is that $\phi_-$ and $\phi_+$ have 
masses $\simeq 750~{\rm GeV}$ and $\simeq 2~{\rm TeV}$, with the 
latter responsible for the diboson excess reported by the ATLAS 
collaboration~\cite{Aad:2015owa} through the $WW$ and $ZZ$ 
decay modes~\cite{Cacciapaglia:2015nga}.
\begin{figure}[t]
\centering
  \includegraphics[clip,width=.49\textwidth]{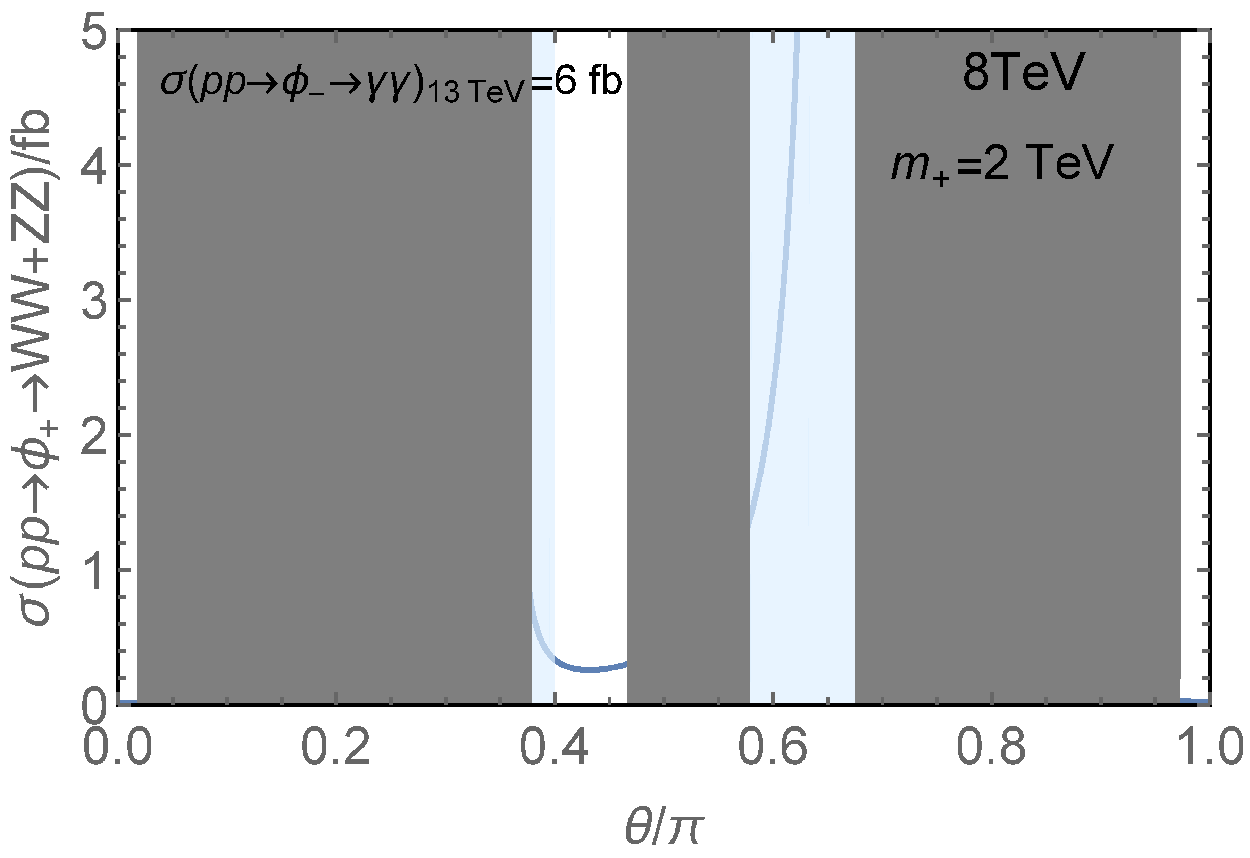}
\caption{The production cross section of $\phi_+$ of mass $m_+ = 
 2~{\rm TeV}$ times the sum of the branching ratios into $WW$ and 
 $ZZ$ at $\sqrt{s} = 8~{\rm TeV}$.}
\label{fig:phi+_2000}
\end{figure}
In Fig.~\ref{fig:phi+_2000}, we show the production cross section of $WW$ 
and $ZZ$ through $\phi_+$.  The meaning of the shades is the same as in 
Fig.~\ref{fig:phi-}.  We find that the cross section of a few fb, which 
is needed to explain the excess, may be achieved for $\theta/\pi\sim 0.6$, 
although chiral perturbation theory can give only qualitative estimates 
in this region.
\end{itemize}
We note that in addition to $\phi_-$ and $\phi_+$, the model also has 
$\eta'$ associated with the anomalous $U(1)$ axial flavor symmetry.  If 
its mass is somewhat lower than the naive expectation and if the $G_H$ 
sector respects $CP$, then we have three scalar resonances decaying into 
two standard model gauge bosons in an interesting mass range.  In this 
case, we may consider even richer possibilities; for example, $\phi_\pm$ 
may be responsible for the $750~{\rm GeV}$ excess with an apparent wide 
width, while $\eta'$ may be responsible for the $2~{\rm TeV}$ diboson 
excess.

Collider phenomenology of $\psi$, $\chi$, and $\varphi$, is essentially 
the same as that in the model without the singlet, except that if there 
are couplings of the form $h^\dagger \Psi_L \bar{\Psi}_N$ and $h \Psi_N 
\bar{\Psi}_L$, giving a (small) mixing between $\lambda$ and the Higgs 
boson, then $\varphi$ may decay into two Higgs bosons through mixing 
involving $\lambda$.  The $\xi$ particle is pair produced by ordinary 
QCD processes.  Its signals depend on the lifetime, and are similar 
to those of the charge $\pm 1/3$ component of $\chi$.  The $\lambda$ 
particle is pair produced by electroweak processes.  Signals again 
depend on the lifetime.  Suppose it is stable at collider timescales. 
The electrically charged component of $\lambda$ is heavier than the 
neutral component by about $360~{\rm MeV}$.  The charged component 
thus decays into the neutral component and a charged (standard model) 
pion with a decay length of $\sim 10~{\rm mm}$, which might be observed 
as a disappearing track.  Since the decay length is short, however, the 
current LHC data do not constrain the mass of $\lambda$.  If $\lambda$ 
decays promptly, then it can be observed as a resonance.  Since $\lambda$ 
has the same standard model gauge quantum numbers as the standard model 
Higgs doublet, possible decay modes of $\lambda$ resemble those of heavy 
Higgs bosons in two Higgs doublet models.

\subsection{Hidden Quasi-Conserved Quantities and Cosmology}
\label{subsec:SU6-cosmo}

In the model with a singlet hidden quark, approximate symmetries in the 
$G_H$ sector can be more easily broken than in the model without it. 
For this purpose, the mass of the singlet hidden quark need not be smaller 
than $\Lambda$, so the following discussion applies even if the mass of 
the singlet hidden quark is larger than the hidden dynamical scale.

``$L$ number'' can be broken by a renormalizable term
\begin{equation}
  {\cal L} \sim h^\dagger \Psi_L \bar{\Psi}_N 
    + h \Psi_N \bar{\Psi}_L + {\rm h.c.},
\label{eq:SU6-dim4}
\end{equation}
which mixes $\lambda$ with the standard model Higgs field.%
\footnote{If $G_H$ is in a conformal phase, the dimensions of operators 
 $\Psi \bar{\Psi}$ are expected to be smaller than~3.  In this case, if 
 the coefficients of the interactions in Eq.~(\ref{eq:SU6-dim4}) are of 
 order one, these interactions participate in the strong dynamics and 
 affect the phase of the theory.  Since the effect of this is not clear, 
 here we assume that the coefficients are smaller than order unity at 
 all scales.  This requires the coefficients to be much smaller than 
 order one at the scale $M_*$.  This smallness of the coefficients at 
 $M_*$ can be understood by an appropriate chiral symmetry, although 
 this symmetry cannot commute with $SU(5)$ if we want to keep the size 
 of the $D$ number violating interactions in Eq.~(\ref{eq:SU6-dim6}).}
Unless this mixing is significantly suppressed, $\lambda$ decays promptly 
into a pair of standard model particles, with the decay channels similar 
to heavy Higgs bosons in two Higgs doublet models.  Furthermore, if 
$m_\chi > m_\xi$, $\chi$ decays into $\xi$ and a standard model Higgs 
or gauge boson through the emission of (off-shell)~$\lambda$.  ``$D$ 
number'' is broken by operators
\begin{equation}
  {\cal L} \sim 
  \left\{ \begin{array}{l}
     \Psi_D \bar{\Psi}_N 
  \end{array} \right\} \times 
  \left\{ \begin{array}{l}
    u d \\ q l 
     \end{array} \right\}
  + {\rm h.c.}
\label{eq:SU6-dim6}
\end{equation}
These operators induce decay of $\xi$ into a pair of standard model 
fermions.  They also induce decay of $\chi$ into a pair of standard 
model fermions and $\lambda$ (or a standard model Higgs or gauge boson 
through the mixing of Eq.~(\ref{eq:SU6-dim4})).  Since the operator 
in the first bracket of Eq.~(\ref{eq:SU6-dim6}) is scalar, it may be 
significantly enhanced by conformal dynamics of $G_H$.  Note that the 
dimension of the operators in the second bracket of Eq.~(\ref{eq:SU6-dim6}) 
is smaller by one than that of Eq.~(\ref{eq:dim7-1}).  Thus, if the 
$G_H$ sector is in a conformal phase between $M_*$ and $\Lambda$, then 
the required size of anomalous dimensions of $\Psi\bar{\Psi}$ to achieve 
the same decay rate of $\chi$ is smaller by one than in the model 
without the singlet hidden quark.

Hidden baryon number can also be violated by non-renormalizable 
interactions involving the singlet hidden quark.  For $N=3$ and 
$N=4$, the lowest dimension operators violating hidden baryon number 
still have the same dimensions as those in Eqs.~(\ref{eq:B-1_N=3}) and 
(\ref{eq:B-1_N=4}).  For $N=5$, however, the following operators exist:
\begin{equation}
  {\cal L} \sim {\bf 5}^* \Psi_{{\bf 5}^*} \Psi_{{\bf 5}^*} 
      \Psi_{{\bf 5}^*} \Psi_{{\bf 5}^*} \Psi_N 
    + {\bf 10} \Psi_{{\bf 5}^*} \Psi_{{\bf 5}^*} \Psi_N 
      \Psi_N \Psi_N 
    + {\bf 10} \bar{\Psi}_{{\bf 5}^*} \bar{\Psi}_{{\bf 5}^*} 
      \bar{\Psi}_{{\bf 5}^*} \bar{\Psi}_N \bar{\Psi}_N 
    + {\rm h.c.},
\label{eq:SU6-baryon}
\end{equation}
in which the dimension of the standard model operators is smaller by one 
than that in Eq.~(\ref{eq:B-1_N=5}).  Here, to make the expression compact, 
we have combined the hidden quarks and standard model fermions into 
$SU(5)$ multiplets:\ ${\bf 5}^* = \{ d, l \}$, ${\bf 10} = \{ q, u, e \}$, 
$\Psi_{{\bf 5}^*} = \{ \Psi_D, \Psi_L \}$, and $\bar{\Psi}_{{\bf 5}^*} 
= \{ \bar{\Psi}_D, \bar{\Psi}_L \}$.  These operators induce mixings 
between hidden baryons and standard model fermions.  Because of the 
lower dimensionality, the required size of anomalous dimensions of the 
hidden baryonic operators (in the case that $G_H$ is in a conformal 
phase between $M_*$ and $\Lambda$) is smaller by one than in the model 
without the singlet hidden quark.

With superparticles around the TeV scale, approximate symmetries in 
the $G_H$ sector are even more easily broken.  In particular, $D$ and 
$L$ numbers are broken by the following superpotential operators
\begin{equation}
  W \sim H_u \Psi_L \bar{\Psi}_N + H_d \Psi_N \bar{\Psi}_L 
    + {\bf 10}\, {\bf 10}\, \Psi_N \bar{\Psi}_{{\bf 5}^*} 
    + {\bf 10}\, {\bf 5}^* \Psi_{{\bf 5}^*} \bar{\Psi}_N,
\label{eq:SU6-SUSY}
\end{equation}
where $H_u$ and $H_d$ are the up-type and down-type Higgs superfields, 
respectively.  Since these are relatively lower-dimensional operators, 
particles whose stability would be ensured by $D$ and $L$ numbers, i.e.\ 
$\chi$, $\xi$ and $\lambda$, decay with cosmologically short lifetimes. 
For example, with the dimension-five operators suppressed by the scale 
$M_* \lesssim 10^{17}~{\rm GeV}$, these particles decay with a lifetime 
shorter than $O(100~{\rm sec})$ even if $G_H$ is not in a conformal phase 
between $M_*$ and the TeV scale.  With $G_H$ not being in a conformal 
phase, the coefficients of the first two operators need not be suppressed 
much:\ coefficients of $O(0.1)$ or smaller are enough to make the mixing 
between $\lambda$ and Higgs fields sufficiently small to preserve their 
respective phenomenology.  Hidden baryon number can also be easily 
broken.  For $N=3$, there exist dimension-five superpotential operators 
violating hidden baryon number.  With the suppression scale $M_* \lesssim 
10^{17}~{\rm GeV}$, the lifetime of hidden baryons is shorter than 
$O(100~{\rm sec})$ even if $G_H$ is not in a conformal phase between 
$M_*$ and the TeV scale.  Having conformal dynamics even allows for 
the decay before the onset of the big bang nucleosynthesis, although 
the coefficients of the first two operators in Eq.~(\ref{eq:SU6-SUSY}) 
need to be appropriately suppressed in this case.

\section{Discussion}
\label{sec:discuss}

In this paper we have studied a simple model in which the recently reported 
diphoton excess arises from a composite pseudo~Nambu-Goldstone boson, 
produced by gluon fusion and decaying into two photons.  In the minimal 
version, the model only has a new hidden gauge group $G_H$ at the TeV scale 
with a hidden quark in the vectorlike bifundamental representation of $G_H$ 
and $SU(5) \supset G_{\rm SM}$.  We have found that the model predicts 
hidden pions $\psi({\bf Adj}, {\bf 1}, 0)$, $\chi(\Box, \Box, -5/6)$, 
and $\varphi({\bf 1}, {\bf Adj}, 0)$, in addition to the diphoton resonance 
$\phi({\bf 1}, {\bf 1}, 0)$, and that the masses of $\psi$ and $\chi$ 
are smaller than $1.6~{\rm TeV}$ and $1.2~{\rm TeV}$, respectively.  The 
existence of these particles, therefore, can be probed at the LHC in the 
near future.  We have studied physics of would-be stable particles---$\chi$ 
and low-lying hidden baryons---in detail, including constraints from 
cosmology.  We have discussed possible theoretical structures above the 
TeV scale, including conformal dynamics and supersymmetry, and their 
phenomenological implications.

In the extended version of the model, there is an additional hidden 
quark that is singlet under the standard model gauge group and has a 
mass smaller than the hidden dynamical scale.  This yields two hidden 
pions that can be produced by gluon fusion and decay into standard model 
dibosons, including a diphoton.  We have discussed several scenarios 
in which these and other resonances can be used to explain various 
excesses seen in the LHC data.  The existence of the singlet hidden 
quark also helps to write operators inducing decays of would-be stable 
particles, such as $\chi(\Box, \Box, -5/6)$, $\xi(\Box,{\bf 1},-1/3)$, 
$\lambda({\bf 1},\Box,1/2)$ hidden mesons and low-lying hidden baryons. 
In particular, if the theory becomes supersymmetric near the TeV scale, 
the scale suppressing these higher dimensional operators can be as high 
as the unification scale $M_* \simeq 10^{16}~{\rm GeV}$ while avoiding 
all the cosmological bounds.

While we have presented it as that explaining the $750~{\rm GeV}$ 
diphoton excess, the model discussed here may also be used to explain 
other diphoton/diboson excesses that might be seen in future data at 
the LHC or other future colliders.%
\footnote{We could use the model to explain the $2~{\rm TeV}$ diboson 
 excess seen in the $8~{\rm TeV}$ ATLAS data~\cite{Aad:2015owa} by 
 decays of $\phi$ produced by gluon fusion.  This, however, requires 
 $m_\phi \simeq 2~{\rm TeV}$ and $f \simeq 100~{\rm GeV}$, since the 
 $WW + ZZ$ cross section at $8~{\rm TeV}$ is given by $4.7~{\rm fb}\, 
 (N/5)^2 (100~{\rm GeV}/f)^2$.  Therefore, the hidden pion picture 
 is not good, i.e.\ $m_\phi \gtrsim \Lambda$, in this case.}
In particular, our studies of would-be stable particles and the structure 
of theories at higher energies can be applied in much wider contexts. 
We hope that some (if not all) of the analyses in this paper are useful 
in understanding future data from experiments.

\section*{Acknowledgments}

We would like to thank Simon Knapen and Dean Robinson for pointing out 
a factor error in Eq.~(\ref{eq:pion-couplings}) in the earlier version. 
This work was supported in part by the Director, Office of Science, 
Office of High Energy and Nuclear Physics, of the U.S.\ Department 
of Energy under Contract DE-AC02-05CH11231, by the National Science 
Foundation under grants PHY-1316783 and PHY-1521446, and by MEXT 
KAKENHI Grant Number 15H05895.

\appendix

\section{Hidden Sector {\boldmath $CP$}}
\label{sec:CP}

In this appendix, we consider the effects of possible $CP$ violation 
in the $G_H$ sector on the standard model.  We work in the field basis 
in which the $\theta$ parameter of the $G_H$ gauge theory vanishes:\ 
$\theta_H = 0$.  (We take this basis by rotating the phase of $\Psi_L 
\bar{\Psi}_L$, not of $\Psi_D \bar{\Psi}_D$, in order not to affect 
the QCD $\theta$ parameter.)  In this case, all the $CP$ violating 
effects in the $G_H$ sector are encoded in the phases of the hidden 
quark masses, which we denote with the capital letters, $M_D$ and 
$M_L$, to remind ourselves that they are in general complex.

We first analyze the effect on the QCD $\theta$ parameter.  For this 
purpose, we consider a nonlinearly realized field
\begin{equation}
  U(x) = \exp\! \left[ \frac{2i}{f} \sum_{A=1}^{24} \xi^A(x) t^A \right],
\label{eq:app-U}
\end{equation}
where $t^A$ are the generators of $SU(5)$, normalized such that 
${\rm tr}[t^A t^B] = \delta^{AB}/2$, and $\xi^A(x)$ are canonically 
normalized hidden pion fields.  The relevant part of the Lagrangian 
is then given by
\begin{equation}
  {\cal L} = -\frac{f^2}{4} {\rm tr}
    \bigl[ ({\cal D}_\mu U) ({\cal D}^\mu U)^\dagger \bigr] 
    + \frac{c}{2} {\rm tr} [M U^\dagger + M^\dagger U] 
    - \frac{i N g_3^2}{128\pi^2} {\rm tr} \bigl[ 
      \bigl(\{ t^a,t^b \} + \{ t^{a*},t^{b*} \}\bigr) \ln U \bigr] 
      \epsilon^{\mu\nu\rho\sigma} G^a_{\mu\nu} G^b_{\rho\sigma},
\label{eq:app-L}
\end{equation}
where the first term is the kinetic term, the second term arises from 
the hidden quark masses
\begin{equation}
  M = \left( \begin{array}{cc}
    M_D {\bf 1}_{3 \times 3} & 0 \\
    0 & M_L {\bf 1}_{2 \times 2}
  \end{array} \right),
\label{eq:app-M}
\end{equation}
and the third term is determined by chiral anomalies in which we have 
kept only the gluon field, where $t^a$ ($a=1,\cdots,8$) are the generators 
of $SU(3) \subset SU(5)$ corresponding to the standard model color. 
In our field basis, $c > 0$.

The Lagrangian of Eq.~(\ref{eq:app-L}) induces a vacuum expectation 
value of the $\xi^{24}$ field, which corresponds to the hidden pion 
$\phi$:\ $\xi^{24} = \langle \xi^{24} \rangle + \phi$.  The relevant 
terms in Eq.~(\ref{eq:app-L}) are
\begin{equation}
  {\cal L} \supset -\frac{1}{2} \partial_\mu \xi^{24} \partial^\mu \xi^{24} 
    - V(\xi^{24}) 
    + \frac{N g_3^2}{32\sqrt{15}\pi^2 f} \xi^{24} 
    \epsilon^{\mu\nu\rho\sigma} G^a_{\mu\nu} G^a_{\rho\sigma},
\label{eq:app-L-xi24}
\end{equation}
where
\begin{equation}
  V(\xi^{24}) = 
  -3c\, m_D \cos\biggl( - \frac{2}{\sqrt{15}f}\xi^{24} + \theta_D \biggr) 
  -2c\, m_L \cos\biggl( \frac{3}{\sqrt{15}f}\xi^{24} + \theta_L \biggr).
\label{eq:app-V}
\end{equation}
Here, we have defined $M_D = m_D e^{i\theta_D}$ and $M_L = m_L 
e^{i\theta_L}$.  The potential of Eq.~(\ref{eq:app-V}) gives
\begin{equation}
  m_D \sin\biggl( - \frac{2}{\sqrt{15}f} \langle \xi^{24} \rangle 
    + \theta_D \biggr) 
  = m_L \cos\biggl( \frac{3}{\sqrt{15}f} \langle \xi^{24} \rangle 
    + \theta_L \biggr).
\label{eq:app-V-prime}
\end{equation}
The solution for $\langle \xi^{24} \rangle$ takes the form
\begin{equation}
  \frac{2}{\sqrt{15}f} \langle \xi^{24} \rangle
  = \theta_D + g(3\theta_D + 2\theta_L; m_D, m_L),
\label{eq:app-xi-vev}
\end{equation}
where $g(x; m_D, m_L)$ is the solution to
\begin{equation}
  m_D \sin(-g) = m_L \sin\biggl( \frac{3}{2}g + \frac{x}{2} \biggr),
\label{eq:app-g}
\end{equation}
which can be expanded for small $x$ as $g(x; m_D, m_L) = 
-(m_L/(2 m_D + 3 m_L)) x + O(x^2)$.  If $m_D = m_L = 0$, $\langle 
\xi^{24} \rangle$ is undetermined at this level, and $\xi^{24}$ 
would act as the QCD axion.

Through the last term in Eq.~(\ref{eq:app-L-xi24}), a nonzero value of 
$\langle \xi^{24} \rangle$ generates a contribution to the QCD $\theta$ 
parameter
\begin{equation}
  \varDelta\theta = -\frac{2N}{\sqrt{15} f} \langle \xi^{24} \rangle 
  = -N \bigl\{ \theta_D + g(3\theta_D+2\theta_L-\theta_H; m_D, m_L) \bigr\},
\label{eq:app-theta}
\end{equation}
where we have restored $\theta_H$ using the fact that only the quantities 
invariant under the phase rotation of $\Psi_L \bar{\Psi}_L$ can 
appear here.  This expression, therefore, applies in any field basis. 
In the lack of a low-energy adjustment mechanism such as a QCD axion 
(and barring accidental cancellation), this contribution must be tiny, 
$|\varDelta\theta| \lesssim 10^{-9}$.  Aside from the possibility of 
fine-tuning between the first and second terms in Eq.~(\ref{eq:app-theta}), 
this requires
\begin{equation}
  \left| \frac{2m_D\theta_D-2m_L\theta_L+m_L\theta_H}{2m_D+3m_L} \right| 
  \lesssim 10^{-9} \frac{1}{N}.
\label{eq:app-limit}
\end{equation}
For $m_{D,L} \neq 0$, this generically forces all the physical phases to 
be tiny, rendering it unlikely that the hidden $\eta'$ decays into two 
hidden pions with a significant rate.  We stress that this constraint 
disappears if there is a QCD axion.

In the presence of a QCD axion, the contribution to $\theta$ is harmless. 
We, however, still expect that the presence of $CP$ violation in the $G_H$ 
sector generates the following dimension-six operator
\begin{equation}
  {\cal L} = \frac{1}{\kappa^2}\, d^{abc} \epsilon^{\nu\rho\sigma\lambda} 
    G^a_{\mu\nu} G^b_{\rho\sigma} G^c_{\lambda}{}^\mu.
\label{eq:app-Weinberg}
\end{equation}
This operator induces the neutron electric dipole 
moment of order $d_n \sim 10^{-26}~e\, {\rm cm} \times 
(100~{\rm TeV}/\kappa)^{-2}$~\cite{Weinberg:1989dx}, so the 
scale $\kappa$ must satisfy $\kappa \gtrsim 100~{\rm TeV}$. 
In the present model, we have
\begin{equation}
  \frac{1}{\kappa^2} \sim 
    \frac{g_3^3 N}{16\pi^2} \frac{m}{\Lambda^3} 
    h(3\theta_D+2\theta_L-\theta_H; m_D, m_L),
\label{eq:app-kappa2}
\end{equation}
where $m$ collectively denotes $m_{D,L}$, and $h(x; m_D, m_L)$ is a 
dimensionless function.  In the relevant parameter region of $m \approx 
O(100~{\rm GeV})$, $\Lambda \sim \mbox{a few TeV}$, $N \sim \mbox{a few}$, 
we find $\kappa \sim 100~{\rm TeV}$, so the contribution can be sufficiently 
small and yet may be seen in future experiments.  The $G_H$ sector 
$CP$ violation also contributes generally to the quark/lepton electric 
dipole and chromoelectric dipole operators, ${\cal L} \sim i\bar{\psi}_L 
\sigma_{\mu\nu} \psi_R F^{\mu\nu} + {\rm h.c.}$ and $i\bar{\psi}_L 
\sigma_{\mu\nu} t^a \psi_R G^{a\mu\nu} + {\rm h.c.}$, but these 
contributions are suppressed by an extra loop factor, so they are 
not very constraining.  Overall, if there is a QCD axion, we may consider 
the possibility of significant $CP$ violation in the $G_H$ sector, 
making the hidden $\eta'$ decay dominantly into two hidden pions.

\end{document}